\documentclass[sn-mathphys-num,referee]{sn-jnl}

\usepackage{graphicx}%
\usepackage{multirow}%
\usepackage{amsmath,amssymb,amsfonts}%
\usepackage{amsthm}%
\usepackage{mathrsfs}%
\usepackage[title]{appendix}%
\usepackage{xcolor}%
\usepackage{textcomp}%
\usepackage{manyfoot}%
\usepackage{booktabs}%
\usepackage{natbib}%
\usepackage{algorithm}%
\usepackage{algorithmicx}%
\usepackage{algpseudocode}%
\usepackage{listings}%
\usepackage{siunitx}

\usepackage{bm}
\usepackage{cleveref}
\usepackage{subcaption}

\usepackage{float}

\newcommand{\data}{\bm{D}}
\newcommand{\ftheta}{\bm{\theta}}

\newcommand{\prior}{p(\ftheta)}
\newcommand{\likelihood}{p(\data | \ftheta)}
\newcommand{\likelihoodcal}{\mathcal{L}(\ftheta)}

\newcommand{\posterior}{p(\ftheta | \data)}
\newcommand{\model}{\mathcal{M}}

\newcommand{\changed}[1]{\textcolor{black}{#1}}
\newcommand{\changedIKM}[1]{\textcolor{black}{#1}}

\begin{document}

\title{\changedIKM{Bayesian Updating of constitutive parameters under hybrid uncertainties with a novel surrogate model applied to biofilms}}


\author[1,2]{\fnm{Lukas} \sur{Fritsch}}\email{fritsch@irz.uni-hannover.de}
\author*[3]{\fnm{Hendrik} \sur{Geisler}}\email{geisler@ikm.uni-hannover.de}
\author[4]{\fnm{Jan} \sur{Grashorn}}\email{grashorn@hsu-hh.de}
\author[3]{\fnm{Felix} \sur{Klempt}}\email{klempt@ikm.uni-hannover.de}
\author[3]{\fnm{Meisam} \sur{Soleimani}}\email{soleimani@ikm.uni-hannover.de}
\author[1]{\fnm{Matteo} \sur{Broggi}}\email{broggi@irz.uni-hannover.de}
\author[3,2]{\fnm{Philipp} \sur{Junker}}\email{junker@ikm.uni-hannover.de}
\author[1,2,5,6]{\fnm{Michael} \sur{Beer}}\email{beer@irz.uni-hannover.de}

\affil[1]{\orgdiv{Institute for Risk and Reliability}, \orgname{Leibniz University Hannover}, \orgaddress{\street{Callinstr. 34}, \city{Hanover}, \postcode{30167}, \country{Germany}}}

\affil[2]{\orgdiv{International Research Training Group (IRTG) 2657}, \orgname{Leibniz University Hannover}, \orgaddress{\street{Appelstr. 11/11a}, \city{Hanover}, \postcode{30167}, \country{Germany}}}

\affil[3]{\orgdiv{Institute of Continuum Mechanics}, \orgname{Leibniz University Hannover}, \orgaddress{\street{An der Universität 1}, \city{Garbsen}, \postcode{30823}, \country{Germany}}}

\affil[4]{\orgdiv{Chair of Engineering Materials and Building Preservation}, \orgname{Helmut-Schmidt-University}, \orgaddress{\street{Holstenhofweg 85}, \city{Hamburg}, \postcode{22043}, \country{Germany}}}

\affil[5]{\orgdiv{Department of Civil and Environmental Engineering}, \orgname{University of Liverpool}, \orgaddress{\city{Liverpool}, \postcode{L69 3GH}, \country{UK}}}

\affil[6]{\orgdiv{International Joint Research Center for Resilient Infrastructure \& International Joint Research Center for Engineering Reliability and Stochastic Mechanics}, \orgname{Tongji University}, \orgaddress{\street{1239 Siping road}, \city{Shanghai}, \postcode{200092}, \country{People’s Republic of China}}}


\abstract{
Accurate modeling of bacterial biofilm growth is essential for understanding their complex dynamics in biomedical, environmental, and industrial settings. These dynamics are shaped by a variety of environmental influences, including the presence of antibiotics, nutrient availability, and inter-species interactions, all of which affect species-specific growth rates. However, capturing this behavior in computational models is challenging due to the presence of hybrid uncertainties, a combination of epistemic uncertainty (stemming from incomplete knowledge about model parameters) and aleatory uncertainty (reflecting inherent biological variability and stochastic environmental conditions).
In this work, we present a Bayesian model updating (BMU) framework to calibrate a recently introduced multi-species biofilm growth model. To enable efficient inference in the presence of hybrid uncertainties, we construct a reduced-order model (ROM) derived using the Time-Separated Stochastic Mechanics (TSM) approach. TSM allows for an efficient propagation of aleatory uncertainty, which enables single-loop Bayesian inference, thereby avoiding the computationally expensive nested (double-loop) schemes typically required in hybrid uncertainty quantification.
The BMU framework employs a likelihood function constructed from the mean and variance of stochastic model outputs, enabling robust parameter calibration even under sparse and noisy data. We validate our approach through two case studies: a two-species and a four-species biofilm model. Both demonstrate that our method not only accurately recovers the underlying model parameters but also provides predictive responses consistent with the synthetic data.
}




\maketitle

\section{Introduction}
\label{sec:introduction}

Bacterial biofilms are structured microbial communities whose growth dynamics are influenced by environmental conditions, nutrient availability, antibiotics, and inter-species interactions \cite{donlan2002}. A key feature of biofilms is their remarkable resilience: they can exhibit up to 1000-fold greater tolerance to antibiotics and environmental stressors compared to planktonic (free-floating) bacteria \cite{kang2023}. This inherent robustness contributes to the widespread presence of biofilms across diverse settings, such as natural ecosystems, industrial systems, and clinical environments.
In industrial and environmental contexts, biofilms can play beneficial roles, such as in wastewater treatment processes \cite{chattopadhyay2022}. However, they are also associated with numerous challenges, including persistent infections \cite{klapper2010,shree2023}, medical device contamination \cite{khatoon2018}, and infrastructure biofouling \cite{melo1997}. One particularly important area is oral biofilm formation, which can lead to infections around dental implants \cite{paquette2006,kommerein2017,rath2017,feng2021}.

In many environments, biofilms are composed of multiple microbial species that compete for resources and respond collectively to external cues \cite{moons2009,nadell2009,yang2011}.
Understanding the dynamics of such multispecies biofilms and modeling these systems is critical for any application. Some fundamental interaction principles are outlined by \citet{james1995} and they serve as a theoretical foundation of how the species interact, and also how they do not interact in some scenarios.
Depending on the specific area of application, different aspects are modeled and different parameters are used to describe the physical and chemical processes of the biofilm growth.
\citet{ouidir2022} review different approaches for the modeling of such systems and classify them by application to wastewaster treatment, soil, and  biomedical applications. Specifically, they focus on biofilms in the oral cavity which is important for dental applications, also highlighted in \cite{yang2011,marsh2005}.

A recently proposed continuum model by \citet{klempt2025}, derived from the extended Hamilton principle, captures multi-species biofilm growth by introducing abstract material parameters.
However, these parameters are typically unknown and subject to uncertainty. Combined with the inherent stochasticity of biological processes, this presents a significant challenge for constructing predictive and physically meaningful models. Accurate parameter inference is thus essential to identify underlying biofilm properties from data and enable robust modeling.

Various strategies for calibrating biological models have been proposed. Frequentist approaches, such as those reviewed by \citet{read2020}, rely on statistical tests (e.g., Kolmogorov-Smirnov) to compare model output distributions with data. Other works follow similar strategies \cite{gabor2015,mitra2019,mary-huard2011,shewa2024}. In contrast, Bayesian inference provides a probabilistic framework for parameter estimation that naturally incorporates uncertainty \cite{robert2011,wilkinson2007}. While Bayesian methods offer interpretability and flexibility, their computational cost remains a key limitation. Recent developments, including modern Markov Chain Monte Carlo (MCMC), sequential Monte Carlo, and Approximate Bayesian Computation (ABC), have helped mitigate this burden \cite{robert2011}.

Despite their advantages, parameter calibration is often neglected in biofilm modeling. For example, \citet{shewa2024} and \citet{rittmann2018} note that default literature values are frequently used instead of performing parameter calibration. Recent studies have demonstrated the value of Bayesian inference for biofilm models, including parameter estimation with quorum sensing \cite{taghizadeh2020} and the inference of rheological properties from experimental data \cite{nooranidoost2023}.
In the field of computational mechanics, \citet{willmann2022a} propose a Bayesian framework for the calibration of a model of multi-physics bioflm model introduced in Ref. \cite{willmann2022}. The authors present an approach that can also handle unavoidable uncertainty and reduces the computational cost through the use of a Gaussian process surrogate for the log-likelihood. Further, different discrepancy metrics are introduced and compared in their study.

The problem to estimate parameters under uncertainty with limited experimental data is also present in many other fields of applications.
One such application is the estimation of parameters of constitutive models in computational mechanics. \citet{wollner2025} present a Bayesian inference framework which they apply to the parameter estimation of the hyperelastic Ogden model.
A summary of approaches common in applications in mechanical and civil engineering are given in Ref. \cite{bi2023}. 
There, again, we come back to Bayesian inference which allows for the handling of the unknown parameters through a posterior probability distribution of said parameters conditioned on the observed data.
In engineering applications this is commonly referred to as Bayesian model updating (BMU) and related to the contribution of \citet{beck1998}.

Building on recent advancements in BMU for engineering applications, we extend these methods to the biofilm growth model introduced by \citet{klempt2025}. In such systems, it is essential to account for inherent variability caused by biological randomness and stochastic environmental influences. This intrinsic variability introduces {aleatory uncertainty}, which significantly increases the complexity of stochastic model updating.
For robust and reliable inference, it is crucial to represent both {epistemic} (due to limited knowledge) and {aleatory} (inherent randomness) uncertainties, collectively referred to as {hybrid uncertainty} \cite{kiureghian2009,beer2013}. However, hybrid uncertainty poses major computational challenges in Bayesian inference, where propagating stochastic variability through complex models typically requires expensive double-loop procedures \cite{bi2019a,kitahara2022,bi2023}.
To address this, we employ a reduced-order model (ROM) based on the Time-Separated Stochastic Mechanics (TSM) methodology \cite{geisler2023,geisler2025}. TSM approximates the stochastic dynamics by expanding the model response with respect to the uncertain parameters around their expected values and solving a sequence of deterministic evolution equations for the expansion coefficients. This separation of temporal and stochastic components enables efficient forward simulation under aleatory uncertainty, eliminating the need for repeated sampling in time.

The aim of this work is to calibrate a continuum model of biofilm growth, influenced by nutrients and antibiotics, as introduced in \cite{klempt2025}. We employ a Bayesian updating approach to infer key model parameters under hybrid uncertainties.
We model our unknown parameters as parametric probability-boxes (p-boxes) with unknown mean values and fixed coefficients of variations, in order to model natural variability as aleatory uncertainty.
Then, a TSM-ROM is derived for an efficient propagation of the aleatory uncertainties as to not use a double-loop of full-order model calls.
We construct a likelihood function based on summary statistics, i.e., mean and variance, of model responses.

In this paper, we begin with a review of the theoretical background on Bayesian model updating, the treatment of hybrid uncertainties and TSM in \cref{sec:background}. Subsequently, in \cref{sec:application}, we present the application of TSM-ROM to the biofilm model. Finally, we demonstrate the efficacy of our methodology through two case studies, illustrating its accuracy and computational efficiency in \cref{sec:experiments}. Our results highlight the potential for broader applications in uncertainty-aware modeling of biological systems.

\section{Background: Parameter identification under uncertainty}
\label{sec:background}

\changed{To motivate the later chapters and to give some context of the formulations we will first introduce the notations and concepts that are used. This includes the notion of stochastic simulation, as well as parameter identification under sparse and noisy data with Bayesian updating. Furthermore, the concepts of hybrid uncertainties and how to treat these with TSM are introduced.}

\changed{\subsection{Uncertainty modeling}}

\changed{
In engineering and, in particular, in computational mechanics, mathematical models $\model(\ftheta)$ are used to predict quantities of interest $\bm{y}$ based on physical or empirical laws, which are dependent on parameters $\ftheta$.
In practice, these parameters are often not known precisely. Material properties, for instance, may vary due to microstructural heterogeneity and loading can vary due to environmental factors \cite{stefanou2009}. 
To represent such variability, probabilistic frameworks are used. In the simplest case, the input is modeled as an $M$-dimensional random variable $\bm{\Theta}$ with density $p_{\bm{\Theta}}(\ftheta)$ defined on the space $\mathcal{D}_{\bm{\Theta}} \subset \mathbb{R}^M$.
Often, this description needs to be expanded to stochastic processes or random fields to account for temporal and spatial correlations, as outlined in Ref. \cite{stefanou2009}.
The uncertainty modeled in such a way with random variables, process and fields is typically aleatory, i.e., describing some inherent variability \cite{kiureghian2009}.
On the other hand, uncertainty can also be epistemic, i.e., originating from limited knowledge and is therefore referred to as reducible uncertainty.
}

\changed{
Determining how this uncertainty propagates to the model output, i.e., characterizing $p_{\bm{Y}}(\bm{y})$ or related statistics, defines the \emph{forward uncertainty propagation} problem. The corresponding forward model is therefore given by
\begin{equation}
    \model : \mathcal{D}_{\bm{\Theta}} \rightarrow \mathbb{R}^{N_\mathrm{out}}, 
    \qquad 
    \bm{y} = \model(\ftheta).
    \label{eq:model}
\end{equation}
}

\changed{
However, in many applications the probabilistic description of $\bm{\Theta}$ is only partially known, reflecting the epistemic uncertainty introduced above. When observational data for the output $\bm{y}$ are available, the problem reverses: one seeks to infer the parameter values or distributions that are consistent with these observations. This leads to a (stochastic) \emph{inverse} problem.
}
\changed{
This situation arises in biofilm modeling, where many constitutive parameters cannot be measured directly and experimental data exhibit substantial variability \cite{taghizadeh2020}. A calibration framework capable of consistently incorporating measurement noise, model imperfections, and incomplete prior information is therefore required \cite{lye2021}.
For a detailed discussion on different approaches, both deterministic and stochastic, we refer to Ref. \cite{chamoin2025}.
As we will show in the next section, Bayesian inference provides a suitable framework and forms the basis of our inverse problem.
}

\subsection{Bayesian model updating}\label{sec:bmu}
\changed{
Bayesian model updating (BMU) addresses the stochastic calibration problem by combining prior knowledge about the parameters with observational data to reduce epistemic uncertainty. The Bayesian theorem is used to derive a distribution of the parameters $\ftheta$ conditioned on the data $\data$.
To relate the model predictions $\bm{y} = \model(\ftheta)$ to measured data $\data$, a discrepancy term $\bm{\varepsilon}$ is introduced:
\begin{equation}
    \data= \model(\ftheta) + \bm{\varepsilon}.
    \label{eq:model_error}
\end{equation}
}
This term accounts for measurement noise and model inaccuracies, acknowledging that computational models are idealized approximations of reality \cite{lye2021}. A common assumption is that $\bm{\varepsilon}$ follows a zero-mean Gaussian distribution with covariance $\bm{\Sigma}_{\bm{\varepsilon}}$, i.e., $\bm{\varepsilon} \sim \mathcal{N}(\bm{0}, \bm{\Sigma}_{\bm{\varepsilon}})$.

The BMU approach \cite{beck1998} addresses challenges such as incomplete data, observation noise, and model-form uncertainty by treating the parameters $\ftheta$ as random variables. Prior knowledge about these parameters is encoded in a prior distribution $\prior$.

A central component of Bayesian inference is the {likelihood function} $\likelihoodcal = \likelihood$, which quantifies the probability of observing the data $\data$ for a given parameter realization $\ftheta$. Essentially, the likelihood serves as a stochastic measure of fit between model predictions and observations. Its specific form depends on modeling assumptions for the discrepancy term. Under the Gaussian noise assumption, the likelihood is itself a Gaussian distribution centered at the model output and evaluated at the data \cite{beck1998,lye2021}. Its value increases as the data and model \changed{predictions} align more closely.

Bayes' theorem then combines the prior and likelihood to yield the posterior distribution $\posterior$ over the parameters:
\begin{equation}
    \posterior = \frac{\likelihoodcal \,  \prior}{p(\data)},
    \label{eq:BayesianTheorem}
\end{equation}
where $p(\data) = \int \likelihoodcal\, \prior\, \text{d}\ftheta$ is \changed{referred to as} the model evidence. Since this normalizing constant is independent of a fixed set of observations $\data$ and often intractable, the unnormalized posterior is typically used in practice:
\begin{equation}
    \posterior \propto \likelihoodcal \,  \prior.
    \label{eq:BayesianTheoremUnnormalized}
\end{equation}
Note that in \cref{eq:BayesianTheorem} and \cref{eq:BayesianTheoremUnnormalized} the densities are implicitly also conditioned on the model assumptions.

\subsubsection{Bayesian model updating in the presence of hybrid uncertainties}\label{sec:bmu_hybrid}

The previously introduced Bayesian model updating approach treats unknown parameters as random variables with prior distributions that are refined into posterior distributions using observational data. 
This process only reduces {epistemic uncertainty}\changed{, i.e., reducible uncertainty due to limited knowledge about fixed-but-unknown quantities.}
However, many real-world systems exhibit not only epistemic but also {aleatory uncertainty}: inherent randomness that cannot be reduced through further data collection. 

\begin{figure}[t]
    \centering
    \includegraphics[width=\linewidth]{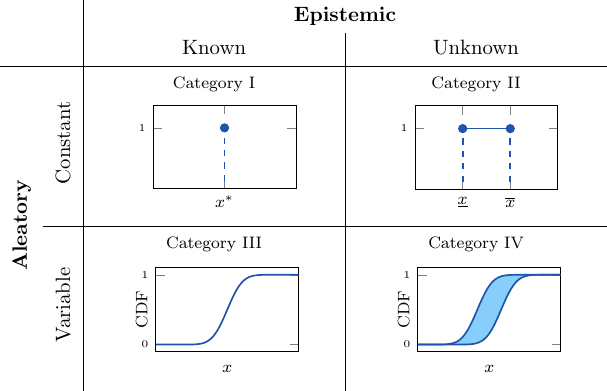}
    \caption{Parameters with different combinations of aleatory and epistemic uncertainties (modified from \cite{dannert2023}).}
    \label{fig:parameters}
\end{figure}

A comprehensive review of model updating under different uncertainty types is provided in \cite{bi2023}, where parameters are categorized into four types based on the presence and combination of aleatory and epistemic uncertainty. An overview of these categories is illustrated in \cref{fig:parameters}. \changed{Category I is defined as fixed and known parameters, the value of a parameter $x$ does not change randomly, i.e. $x$ is not a random variable and is fixed to the value $x^*$. Category II refers to $x$ being bound between $\underline{x}$ and $\overline{x}$, but the true value of $x$ is unknown. Category III then refers to $x$ being randomly distributed according to some known probability distribution, and finally category IV refers to parameters that follow some probability distribution whose parameters are unknown.}

In this work, we focus on the specific challenges posed by parameters of category IV, which involve both types of uncertainty and are thus described by {imprecise probabilities} \cite{beer2013}. A common representation of such parameters is the probability box (p-box), in which aleatory uncertainty is modeled as a random variable, while epistemic uncertainty is expressed as interval bounds on the distribution parameters. \changed{P-boxes are bounded by a lower and upper bound of the cumulative distribution function (CDF), as seen in \cref{fig:parameters}. For more detailed information on p-box modeling we refer to \cite{faes2021}. Because we update the hybrid uncertainties through a Bayesian framework, the outcome is a set of probability distributions for the values of the true model-parameter distributions. In this case we focus on the mean values.}

If models exhibit hybrid uncertainty, their outputs are inherently stochastic. A deterministic mapping like in \cref{eq:model} is no longer sufficient, and commonly a stochastic forward model based on Monte Carlo simulation $\Hat{\model} : \mathcal{D}_{\bm{\Theta}} \rightarrow \mathbb{R}^{N_\mathrm{out} \times N_\mathrm{samples}}$ is defined,
where $N_\mathrm{samples}$ denotes the number of Monte Carlo samples used to propagate aleatory uncertainty for each parameter configuration $\bm{\theta}$. Each input returns a sequence of outputs:
\begin{equation} 
    \mathcal{Y} = \left\{ \bm{y}_1, \bm{y}_2, \dots, \bm{y}_{N_\mathrm{samples}} \right\} = \hat{\model}(\ftheta).
    \label{eq:stochastic_model}
\end{equation}
This setup typically results in a nested (double-loop) approach: an outer loop explores different realizations of $\bm{\theta}$, while the inner loop samples the stochastic model output using the deterministic model $\model$.
Examples for this are the updating approaches in Refs. \cite{bi2019a,kitahara2022,bi2023}.
Although this allows the construction of p-boxes for the model response, it comes at a high computational cost, particularly in Bayesian inverse problems, which require many model evaluations across the parameter space.

In the presence of hybrid uncertainties, constructing the likelihood function becomes especially challenging, since model outputs are no longer deterministic but represent probability distributions.

In simple cases where a single observation $\data$ is available and the model output can be assumed to be Gaussian, the likelihood can be constructed as

\begin{equation}
    \log \likelihoodcal = - \log\left( \sqrt{(2 \pi)^{N_{\text{out}}} \det \bm{\Sigma}_{\bm{\mathcal{Y}}}} \right) - \frac{1}{2}(\data - \bm{\mu}_{\bm{\mathcal{Y}}})^\top \bm{\Sigma}^{-1}_{\bm{\mathcal{Y}}} (\data - \bm{\mu}_{\bm{\mathcal{Y}}}) .
    \label{eq:LogGaussLkl}
\end{equation}
where $\bm{\mu}_{\bm{\mathcal{Y}}}$ and $\bm{\Sigma}_{\bm{\mathcal{Y}}}$ are mean and covariance matrix of the model output. \Cref{eq:LogGaussLkl} uses a logarithmic formulation to reduce issues coming from limited support of the likelihood in the parameter space. Note that for more involved problems with multiple available observations of the same quantity and non-Gaussian model outputs the formulation of \cref{eq:LogGaussLkl} becomes more involved, since the distributions of the model output and the data need to be compared. \changed{An assumption that is made in this paper is that the measurements and model outputs are directly related and can be compared. Note that other setups, e.g. those that use a deterministic model, would require a different setup. For deterministic models usually a measurement error is assumed to formulate a relationship between model outputs and observations, as outline in \cref{eq:model}. Using the model to directly simulate measurements is also known as approximate Bayesian computation \cite{turner2012}.} A detailed discussion of the implications and difficulties of the likelihood construction under hybrid uncertainties is given in appendix \ref{sec:likelihood}. In time-dependent problems, \cref{eq:LogGaussLkl} naturally extends to observations from multiple time instances by taking the sum of the individual measurements (assuming these are independent)

\begin{equation}
    \log \likelihoodcal =  \sum_k - \log\left( \sqrt{(2 \pi)^{N_{\text{out}}} \det \bm{\Sigma}_{\bm{\mathcal{Y}}}^k} \right) - \frac{1}{2}(\data^k - \bm{\mu}_{\bm{\mathcal{Y}}}^k)^\top {\bm{\Sigma}^k_{\bm{\mathcal{Y}},k}}^{-1} (\data^k - \bm{\mu}_{\bm{\mathcal{Y}}}^k) .
    \label{eq:mvLogLkl}
\end{equation}
The likelihood therefore describes the probability of observing the data under some assumption of $\ftheta$. Through the hybrid uncertainties in the model the natural variability of the experiments can be captured and incorporated in the estimates of the model parameters. The specific likelihood implementation for this paper is introduced in \cref{sec:experiments}.

\subsubsection{Transitional Markov Chain Monte Carlo}
In practice, the posterior distribution is analytically intractable, primarily due to its implicit dependence on the forward model through the likelihood function \cite{lye2021}.
Moreover, when multiple measurements are incorporated, the posterior typically does not conform to any standard probability distribution, which becomes clear when considering how the log-likelihood function is constructed in \cref{eq:mvLogLkl}.
To address this, Markov Chain Monte Carlo (MCMC) methods are widely used, as they enable sampling from distributions of arbitrary shape. A key advantage of MCMC techniques is their ability to draw samples directly from the unnormalized posterior distribution, as given in \cref{eq:BayesianTheoremUnnormalized}.

The most common MCMC algorithm is the Metropolis-Hastings algorithm \cite{metropolis1953,hastings1970a} which generates samples from the unnormalized posterior by starting a random walk algorithm with Markov chains \changed{from} some initial samples and than accepting or rejecting new samples based on a proposal distribution.

In this paper, we apply the Transitional Markov Chain Monte Carlo (TMCMC) algorithm \cite{ching2007} to draw samples from \cref{eq:BayesianTheoremUnnormalized}. TMCMC is an advanced MCMC method designed to sample from multimodal distributions by sampling from a sequence of intermediate distributions rather than from the posterior directly, a process which is also known as annealing \cite{kirkpatrick1983}. These transitional densities are defined as
\begin{equation}
     p^j(\ftheta) \propto  \likelihoodcal^{\beta_j} \,  \prior,
     \label{eq:TMCMC}
\end{equation}
where $j \in \{1, 2, \dots, m\}$ represents the transition steps, with the corresponding tempering parameter $\beta_j$ progressing from $\beta_0 = 0$ to $\beta_m = 1$ through intermediate values $\beta_0 = 0 < \beta_1 < \beta_2 < \cdots \beta_m = 1$.
This gradual transition allows the prior density $\prior$ to smoothly evolve into the posterior density $\likelihoodcal \prior$, as noted in Ref. \cite{lye2021}.
To make use of the formulation of the log-likelihood in \cref{eq:mvLogLkl}, the transitional densities in \cref{eq:TMCMC} are formulated \changed{using the logarithmic densities:}
\begin{equation}
    \changed{\log  p^j(\ftheta) \propto  \log \likelihoodcal \cdot {\beta_j} + \log \prior.}
    \label{eq:TMCMC-log}
\end{equation}
We chose TMCMC for its robustness against posteriors with small support due to the annealing procedure.

\subsection{Time-separated Stochastic Mechanics}
In order to reduce the high computational cost of the double-loop approach, a surrogate model or ROM can be used to replace the inner loop and thus decrease the computational cost, as also mentioned in \cite{bi2023}.
\citet{faes2021} discuss different surrogate modeling techniques for the propagation of hybrid uncertainties modeled with p-boxes.
Further, \citet{reiser2025} present a two-step Bayesian framework for surrogate-based inference that  propagates both epistemic and aleatoric uncertainties from surrogate model training to parameter inference.

Here, we use the Time-separated Stochastic Mechanics (TSM) \cite{geisler2023, geisler2025} to efficiently handle the intrinsic aleatory uncertainty and reduce the updating from a double to a single-loop algorithm. 
The main idea of the TSM is to replace a forward model $\model(\ftheta)$ with the parameters $\ftheta$ by a surrogate $\model_S$.  
The surrogate is defined such, that the first $p$ partial derivatives of the surrogate coincide with the forward model at the expectation of the parameters. For a scalar parameter $\theta$ with expectation $\langle \theta \rangle$ this reads as
\begin{equation}
    \frac{\partial^i}{\partial \theta^i} \model \Big|_{\langle \theta \rangle} = \frac{\partial^i}{\partial \theta^i} \model_S \Big|_{\langle \theta \rangle} \quad \forall i \in \{0, \dots, p \}.
\end{equation}
For an algebraic model, this coincides with a Taylor series. Forward models involving derivatives in time need a special treatment. This is presented in more detail in Section~\ref{sec:TSMBiofilm}.
The approach is advantageous for the application to Bayesian Model Updating as the approximation is best near to the expectation of the parameters. Often, much of the probability mass is indeed collected around the expectation. In comparison to other surrogate models, as the Polynomial Chaos Expansion and Stochastic Collocation Method, only a very limited number of function evaluations \changed{are} needed. In fact, for many problems a linear or quadratic approximation suffices \changed{\cite{geisler2025}}.

\section{Application to biofilm growth}
\label{sec:application}

\changed{The proposed Bayesian model updating with the TSM-ROM is applied for infer the constitutive parameters of a biofilm growth model introduced in Ref. \cite{klempt2025}}. This model incorporates the growth of multiple species under the influence of nutrients and antibiotics, as well as their interactions. The growth is represented by the concentration of the biofilms over time, denoted by $\bm{\phi}$. Additionally, $\bm{\psi}$ is defined as the percentage of living bacteria. The volume occupied by living bacteria from species $l$ is expressed as $\overline{\phi}_l = \phi_l \psi_l$.
The model derives from the extended Hamilton principle \cite{junker2021}, which leads for the special case of a local, quasi-static, isothermal model with no external forces to
\begin{equation}\label{eq:localHam}
    \frac{\partial \Psi}{\partial \bm{\xi}} + \frac{\partial \Delta^s}{\partial \Dot{\bm{\xi}}} + \frac{\partial c}{\partial \bm{\xi}} = 0,
\end{equation}
with the vector of internal variables 
\begin{equation}
    \bm{\xi} = \begin{pmatrix} \bm{\phi} \\ \bm{\psi} \end{pmatrix},
\end{equation}
the energy density function $\Psi$, the dissipation function $\Delta^s$ and the constraint function $c$.

The temporal evolution of biofilm concentration is determined by the energy density function and the dissipation function. The energy density function is defined as
\begin{equation}
\Psi = - \frac{1}{2} \mathrm{c}^* \overline{\bm{\phi}} \cdot \boldsymbol{A} \cdot \overline{\boldsymbol{\phi}} + \frac{1}{2} \alpha^{\star} \boldsymbol{\psi} \cdot \boldsymbol{B} \cdot \boldsymbol{\psi}
\label{eq:biofilm}
\end{equation}
and consists of two terms. The first term, where $\mathrm{c}^*$ represents \changedIKM{the local concentration of nutrients}, which promote an increase in living bacteria, while the second term, where $\alpha^{\star}$ signifies \changedIKM{ the local concentration of antibiotics}, results in a decrease.

Coefficient matrices $\boldsymbol{A}$ and $\boldsymbol{B}$ are crucial in characterizing the material behavior of the species. The matrix $\boldsymbol{A}$ is a symmetric matrix designed to capture the interactions between different biofilm species and the effects of nutrients on their growth. Its off-diagonal elements represent inter-species interactions, while the diagonal elements account for the effects of nutrients on individual species growth. Generally, $\boldsymbol{A}$ can be expressed for multiple species as
\begin{equation}
\boldsymbol{A} =  \begin{pmatrix}
a_{11} & a_{12} & \cdots & a_{1n} \\
a_{12} & a_{22} & \cdots & a_{2n} \\
\vdots & \vdots & \ddots & \vdots \\
a_{1n} & a_{2n} & \cdots & a_{nn}
\end{pmatrix}.
\label{eq:A_matrix}
\end{equation}

The matrix $\boldsymbol{B}$ is a diagonal matrix that characterizes the impact of antibiotics on the viability of the biofilm species. Its diagonal components represent the susceptibility of each species to antibiotics. The general form of $\boldsymbol{B}$ for multiple species is assumed to be
\begin{equation}
\boldsymbol{B} = \begin{pmatrix}
b_{1} & 0 & \cdots & 0 \\
0 & b_{2} & \cdots & 0 \\
\vdots & \vdots & \ddots & \vdots \\
0 & 0 & \cdots & b_{n}
\end{pmatrix}.
\label{eq:B_matrix}
\end{equation}

The dissipation function is modeled as a function of $\Dot{\Bar{\boldsymbol{\phi}}}$ and $\Dot{\bm{\phi}}$ as
\begin{equation}
    \Delta^s = \Delta^s(\Dot{\Bar{\boldsymbol{\phi}}}, \Dot{\boldsymbol{\phi}}) = \frac{1}{2} \; \Dot{\Bar{\boldsymbol{\phi}}} \cdot \boldsymbol{\eta} \cdot \Dot{\Bar{\boldsymbol{\phi}}} + \frac{1}{2} \; \Dot{\boldsymbol{\phi}} \cdot \boldsymbol{\eta} \cdot \Dot{\boldsymbol{\phi}}
    \label{eq:dissipation}
\end{equation}
with the \changedIKM{diagonal matrix representing the rate sensitivity for each species}
\begin{equation}
\boldsymbol{\eta} =
\begin{pmatrix}
\eta_1 & 0 & \cdots & 0\\
0 & \eta_2 & \cdots & 0\\
\vdots & \vdots  & \ddots & \vdots\\
0 & 0 & \cdots & \eta_n
\end{pmatrix}.
\end{equation}
\changedIKM{The dissipation was modeled as a function of the rate of living bacteria $\Dot{\Bar{\boldsymbol{\phi}}}$ and the rate of biofilm concentration $\Dot{\bm{\phi}}$. This modeling approach is based on the assumption that energy is dissipated under two main conditions. The first condition is a change in biofilm concentration due to growth or decay. The second one is not a change in the percentage in living cells but rather a change in the amount of living cells. Since a change of percentage of living cells can not happen without the change in biofilm concentration this derived parameter for the modeling of the dissipation function instead of the state variable $\Dot{\bm{\psi}}$. This leads to a deeply linked system of equation, resulting in complex model behavior.}

To limited the growth to a finite amount, the total volume of all species is limited by the constraint function
\begin{equation}
    c = \gamma \left( \sum^{n}_{l=0}\phi_l-1 \right) = 0
\label{eq:constraint}
\end{equation}
with the Lagrange multiplicator $\gamma$.
\changedIKM{
Inserting Equations \ref{eq:biofilm}, \ref{eq:dissipation} and \ref{eq:constraint} in Equation \ref{eq:localHam} leads to the governing evolution equations in their strong forms:}
\begin{align}
    0 &= -  c^{\star} \psi_i \left( a_{ii} \Bar{\phi}_i+ \sum_j^{n-1} a_{ij} \Bar{\phi}_j \right) + \eta_i (\Dot{\phi_i}\psi_i^2 +\Bar{\phi}_i \Dot{\psi}_i + \Dot{\phi}_i) + \gamma \label{eq:EvolEq1}\\    
    0 &= -c^{\star} \phi_i \left( a_{ii} \Bar{\phi}_i+ \sum_j^{n-1} a_{ij} \Bar{\phi}_j \right) +  \alpha^{\star} \psi_i b_i + \eta_i (\Dot{\psi_i}\phi_i^2 + \Bar{\phi}_i \Dot{\phi}_i) + \gamma \label{eq:EvolEq2}\\
    0 &= \sum_{l=0}^n \phi_l - 1 \label{eq:EvolLagrange}
\end{align}

\changedIKM{A more thorough derivation of the material model can be found in \cite{klempt2025}.}

In the following, the parameters in $\bm{A}$ and $\bm{B}$, which describe the behavior of the species are referred to as the unknown, stochastic parameters $\ftheta$. The observations $\data$ are the volume percentage for each species $\phi_l$ and the percentage of living cells for each species $\psi_l$ for each time step.

To ensure, that the internal variables remained in the interval $\xi_i \in [0,1]$, the penalty method was applied. The penalty terms \changedIKM{$K_p (\frac{1}{\xi_1^2(1-\xi_1)^2}) [\frac{\mathrm{J}}{\mathrm{m}^3}]$} were added for all internal variables to the free energy density in \autoref{eq:biofilm}.


\subsection{Uncertainty model for biofilm growth}\label{sec:uq-model}
\changed{The biofilm-growth parameters $\ftheta$ are modeled as Category IV uncertainties using p-boxes, capturing both aleatory variability and epistemic uncertainty. 
As introduced in \cref{sec:bmu_hybrid}, the resulting stochastic model (see \cref{eq:stochastic_model}) produces a distribution of outputs even for fixed inputs. To efficiently handle this hybrid uncertainty in the inverse problem, we replace the Monte Carlo estimator with a ROM based on TSM, which returns output trajectories for given $\ftheta$. Epistemic uncertainty is encoded in the unknown mean values of $\ftheta$; the likelihood is formulated in terms of these means, on which we place a prior and update it in a Bayesian framework.}

\subsection{TSM-ROM for biofilm growth}
\label{sec:TSMBiofilm}
Equations~\eqref{eq:EvolEq1}, \eqref{eq:EvolEq2} and \eqref{eq:EvolLagrange} constitute a nonlinear system of equations for the evolution of the variables $\bm{\psi}$ and $\bm{\phi}$. 
The input parameter $\ftheta$ is split into an part with epistemic uncertainty $\ftheta^{(0)}$ and a part with aleatoric uncertainties $\tilde{\ftheta}$ with the expectation $\langle \tilde{\ftheta} \rangle = \bm{0}$.
In the following, the input parameters are referred to as $\bm{\theta}^\textrm{TSM} = \bm{\theta}^{(0)} + \tilde{\bm{\theta}}$.
The variables $\bm{\psi}$ and $\bm{\phi}$ are expressed by a linear Taylor series in the aleatoric uncertainties
\begin{align}
    \bm{\psi}^\textrm{TSM} &= \bm{\psi}^{(0)} + \tilde{\ftheta} \cdot \bm{\psi}^{(1)}, \label{eq:PsiTSM} \\
    \bm{\phi}^\textrm{TSM} &= \bm{\phi}^{(0)} + \tilde{\ftheta} \cdot \bm{\phi}^{(1)}. \label{eq:PhiTSM}
\end{align}
The Taylor series for the variables $\bm{\psi}$ and $\bm{\phi}$ constitute the surrogate model $\model_S$. No solution of the systems of equations is needed for a new parameter value $\ftheta$. While a higher order model is simple to implement, a linear series already often suffices \cite{geisler2025}.
For simplicity, let us refer to the system of equations \eqref{eq:EvolEq1} to \eqref{eq:EvolLagrange} as $\mathcal{G}(\bm{\dot{\phi}}, \bm{\dot{\psi}}, \bm{\phi}, \bm{\psi}, \ftheta^\textrm{TSM})$. The system of equations naturally depends on both, the variables themselves and their time derivatives.
The time derivative of the Taylor series results trivially as
\begin{align}
    \dot{\bm{\psi}}^\textrm{TSM} &= \dot{\bm{\psi}}^{(0)} + \tilde{\ftheta} \cdot \dot{\bm{\psi}}^{(1)}, \\
    \dot{\bm{\phi}}^\textrm{TSM} &= \dot{\bm{\phi}}^{(0)} + \tilde{\ftheta} \cdot \dot{\bm{\phi}}^{(1)}.
\end{align}
because the random term $\tilde{\ftheta}$ is time-independent.
All Taylor series are set into the system of equations.
The zeroth order terms can be calculated as
\begin{equation}
    \mathcal{G}(\bm{\dot{\phi}}^\textrm{TSM}, \bm{\dot{\psi}}^\textrm{TSM}, \bm{\phi}^\textrm{TSM}, \bm{\psi}^\textrm{TSM}, \ftheta^\textrm{TSM}) \big\vert_{\tilde{\ftheta} = \bm{0}}. \label{eq:ZerothOrder}
\end{equation}
Unsurprisingly, the original system of equations evaluated at $\ftheta^\textrm{TSM} = \bm{0}$ results.
For the derivation of the first order term,
\begin{equation}
    \frac{\mathrm{d}}{\mathrm{d} \tilde{\ftheta} } \mathcal{G}(\bm{\dot{\phi}}^\textrm{TSM}, \bm{\dot{\psi}}^\textrm{TSM}, \bm{\phi}^\textrm{TSM}, \bm{\psi}^\textrm{TSM}, \ftheta^\textrm{TSM}) \big\vert_{\tilde{\ftheta} = \bm{0}}. \label{eq:FirstOrder}
\end{equation}
is calculated. This results in a nonlinear system of equations that allow to compute $\dot{\bm{\phi}}^{(1)}$ and $\dot{\bm{\psi}}^{(1)}$. \changedIKM{For the solution a Newton-Raphson scheme is used at each time step.} As these equations depend on $\dot{\bm{\phi}}^{(0)}$ and $\dot{\bm{\psi}}^{(0)}$, a staggered solution scheme naturally arises.
\changedIKM{The total algorithm is summarized in Algorithm~\ref{alg:TSMROM}.}

\begin{algorithm}
    \caption{TSM-ROM algorithm}
    \label{alg:TSMROM}
    \begin{algorithmic}
        \Require Input parameter $\ftheta$, Initial state of $\Psi$ and $\Phi$
        \State{Compute expectation of input parameters $\ftheta^{(0)}$}
        \For {each timestep}
        \While {Newton scheme is not converged}
        \State{Evaluate zeroth order Eq.\eqref{eq:ZerothOrder} to get $\bm{\phi}^{(0)}$, $\bm{\Psi}^{(0)}$}
        \EndWhile
        \While {Newton scheme is not converged} 
        \State{Evaluate first order Eq. \eqref{eq:FirstOrder} to get $\bm{\phi}^{(1)}$, $\bm{\Psi}^{(1)}$} 
        \EndWhile
        \EndFor
        \State{Evaluate Eq.\eqref{eq:PsiTSM} and \eqref{eq:PhiTSM} for all input parameters}
    \end{algorithmic}
\end{algorithm}

\section{Numerical Experiments}
\label{sec:experiments}
To illustrate the application of BMU in determining the \textit{material properties} governing biofilm growth, we present two case studies. The primary objective in both cases is to infer the matrices $\boldsymbol{A}$ and $\boldsymbol{B}$, which characterize biofilm evolution through the energy density function outlined in \cref{eq:biofilm}.

\changedIKM{The forward model was solved in an implicit framework on a single material point. To evaluate the governing equations \ref{eq:EvolEq1}, \ref{eq:EvolEq2} and \ref{eq:EvolLagrange} a Newton-Raphson scheme was used at each timestep. This setup is computationally highly efficient. The solution times for the forward model for 2 or 4 species are almost negligible}\changed{, as noted in Section~\ref{sec:CompEff}.}

The first case study involves a model with two species and five unknown material parameters. The second case study extends this framework to four species with 14 parameters. In both scenarios, the BMU approach utilizes TMCMC, implemented via the Julia package \textsf{UncertaintyQuantification.jl} \cite{behrensdorf2025}. To compare model predictions with experimental data, we employ the likelihood function given in \cref{eq:mvLogLkl}, assessing discrepancies in the volume fractions of living bacteria, denoted by $\overline{\bm{\Phi}}$, at predetermined discrete time steps.

To build the likelihood function, the response of the TSM biofilm model is compared to experimental data. 
Specifically, we consider the mean response feature of the $l$-th species at the $k$-th time step, denoted as $\mu_{\bm{y},l}^{k}$. Similarly, its variance is denoted as $\sigma_{\bm{y},l}^{k}$.
In this context, the output features are defined as the volume fractions of $n$ living bacterial species at various discrete time steps $k$. For a single sample $\ftheta_j$, i.e. one full simulation of the TSM-ROM model, these features are structured in a matrix form:

\begin{equation}
    \bm{y}_j = \begin{pmatrix} 
    \overline{\phi}_1^{1} & \overline{\phi}_1^{2} & \cdots & \overline{\phi}_1^{m} \\ 
    \overline{\phi}_2^{1} & \overline{\phi}_2^{2} & \cdots & \overline{\phi}_2^{m} \\ 
    \vdots & \vdots & \ddots & \vdots \\ 
    \overline{\phi}_n^{1} & \overline{\phi}_n^{2} & \cdots & \overline{\phi}_n^{m}
    \end{pmatrix},
    \label{eq:features}
\end{equation}
where $\overline{\phi}_l^{k}$ denotes the volume fraction of species $l$ at time step $k$. Here, $n$ represents the total number of species, while $m$ is the number of selected discrete time steps used for comparison with observational data. For better readability, we omitted the index $j$ in $\overline{\phi}_l^{k}$.

It is important to note that $m \ll N$, where $N$ denotes the total number of time steps in the complete analysis. \changedIKM{For the number of time steps, we take a large number as $N=1000$ in the numerical experiments. This corresponds to a small time step in the order of \SI{e-4}{s}. This is beneficial to the numerical stability and accuracy. The maximal time step is limited by the behavior as the empty phase approaches zero. In this work we chose the time step heuristically by identifying a stable time step for all parameter values. No additional error control is used. }

While $N$ allows for detailed modeling of the biofilm dynamics over time, $m$ is strategically chosen to facilitate the construction of the likelihood function by providing a manageable subset of time steps. For instance, one might select $m=20$ time steps for comparing simulation responses with experimental data. This selection ensures efficient and meaningful statistical inference without overwhelming computational resources.
Thus, a dataset $\data$ is constructed from $N_{\mathrm{data}}=m$ unique experiments, each representing a realization that terminates at different time steps.
Our aim is to replicate the setup of \textit{in vitro} experiments in our \textit{in silico} experiments, i.e., simulating multiple experiments started from the same initial conditions, but stopped at different time instances. 
This approach accounts for both aleatory uncertainty, arising from inherent randomness in biofilm growth, and epistemic uncertainty due to limited observational data. 

Finally, the likelihood function from \cref{eq:mvLogLkl} is computed for each row within the feature matrix detailed in \cref{eq:features}, as described above. In preliminary testing we found that using the full covariance matrix is unnecessary, as correlations between the model outputs are implicitly handled by the model. Therefore, the likelihood is reduced to only the diagonal elements of the covariance matrix, and the expression in \cref{eq:LogGaussLkl} can be simplified. The full likelihood, constructed from the $l \in \{1,\dots, n\}$ species and $k \in \{1,\dots,m\}$ time steps, thus reads

\begin{equation}
    \log \likelihoodcal = \sum_{l} \sum_{k} - \log\left[2 \pi {\sigma_{\bm{y},l}^{k}}^2\right] - \frac{1}{2}{\sigma_{\bm{y},l}^{k}}^{-2} \left(\overline{\phi}_l^{k} - \mu_{\bm{y},l}^{k}\right)^2  .
    \label{eq:mvLogLkl_Biofilm}
\end{equation}
where $\sigma_{\bm{y},l}^{k}$ is the $i$-th diagonal element of $\bm{\Sigma}_{\bm{\mathcal{Y}}}^k$ at time step $t_k$.
Further comment will be given below.

\subsection{Case I: Two-Species Biofilm Model}

In this first case study, we focus on biofilm growth comprising $n=2$ interacting species. Specifically, this case aligns with case 5 in Ref. \cite{klempt2025}, augmented by the addition of an interaction parameter $a_{12}$ to adeptly capture inter-species interactions.

\begin{figure}[t]
    \centering
    \includegraphics[width=\linewidth]{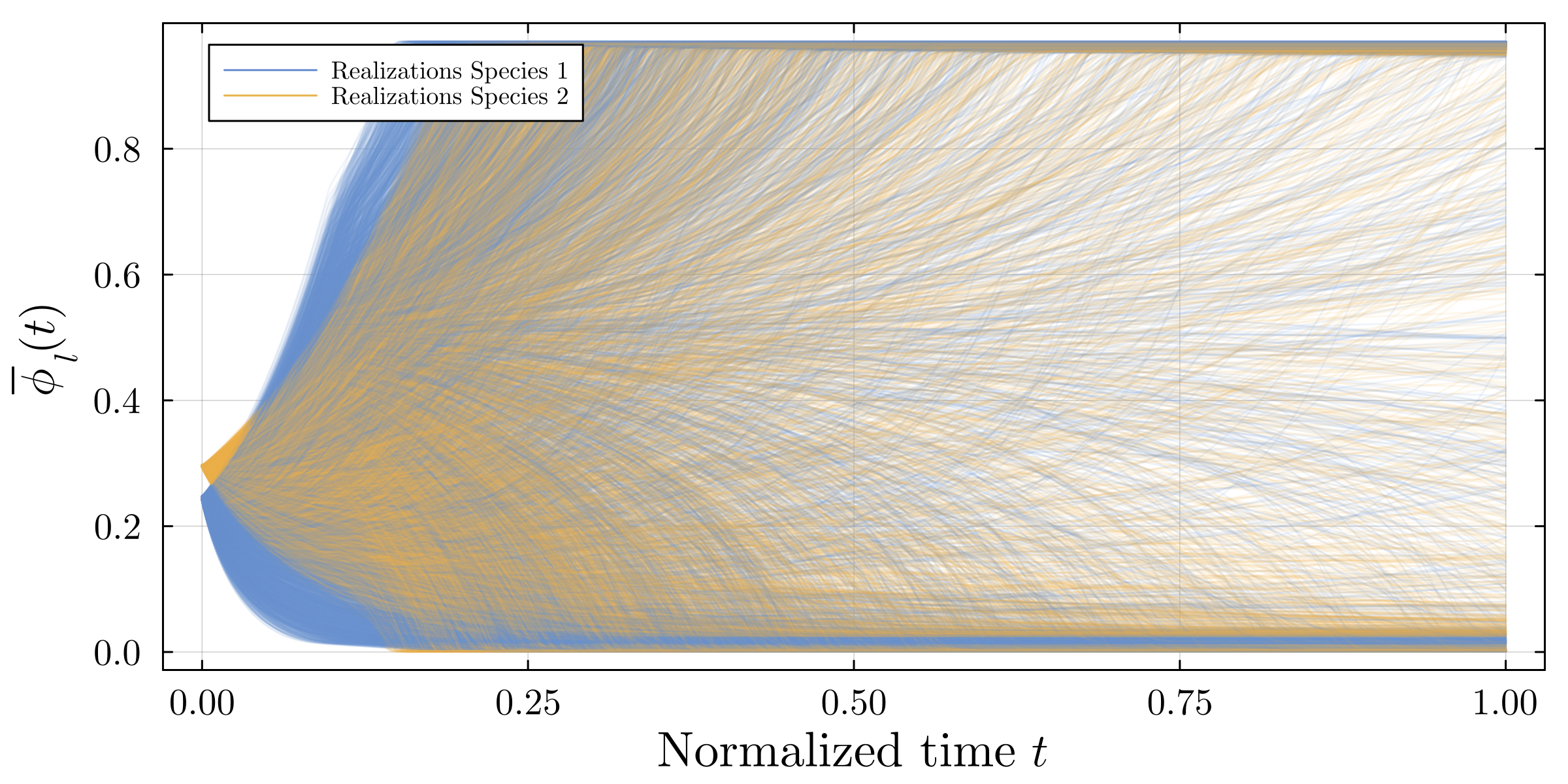}
    \caption{Model realizations corresponding to the input given by the prior samples of case I from \cref{tab:priors-case-I}.}
    \label{fig:prior-output-I}
\end{figure}

For the two-species scenario, the material behavior of the biofilm is characterized by the two matrices in \cref{eq:A_matrix,eq:B_matrix} with $n=2$.
The symmetry inherent in matrix $\boldsymbol{A}$ reduces the dimensionality of the problem to five independent parameters $\ftheta = [a_{11}, \ a_{12}, \  a_{22}, \  b_{1}, \  b_{2}]$. Each of the five parameters is modeled as a parametric p-box with a normal distribution with an unknown mean and a coefficient of variation. \changed{The coefficient of variation controls how much aleatory uncertainty is introduced in the model during the updating process. To show the robustness, two different values are selected, $\mathrm{CoV}^a = 0.5 \%$ and $\mathrm{CoV}^b = 2.0 \%$.}
\changed{For the unknown mean values, uninformative priors modeled by uniform distributions are chosen. The selected ranges of the priors are described in \cref{tab:priors-case-I}.
Samples of the model response corresponding to inputs from the prior are shown in \cref{fig:prior-output-I} to visualize the wide range of possible model outputs when assuming the chosen priors.
Further, we compare the samples obtained from updating with the likelihood in \cref{eq:mvLogLkl_Biofilm} with updating with a likelihood that uses the full covariance, as outlined initially in \cref{eq:mvLogLkl}
.}

\begin{figure}[t]
    \centering
    \includegraphics[width=\linewidth]{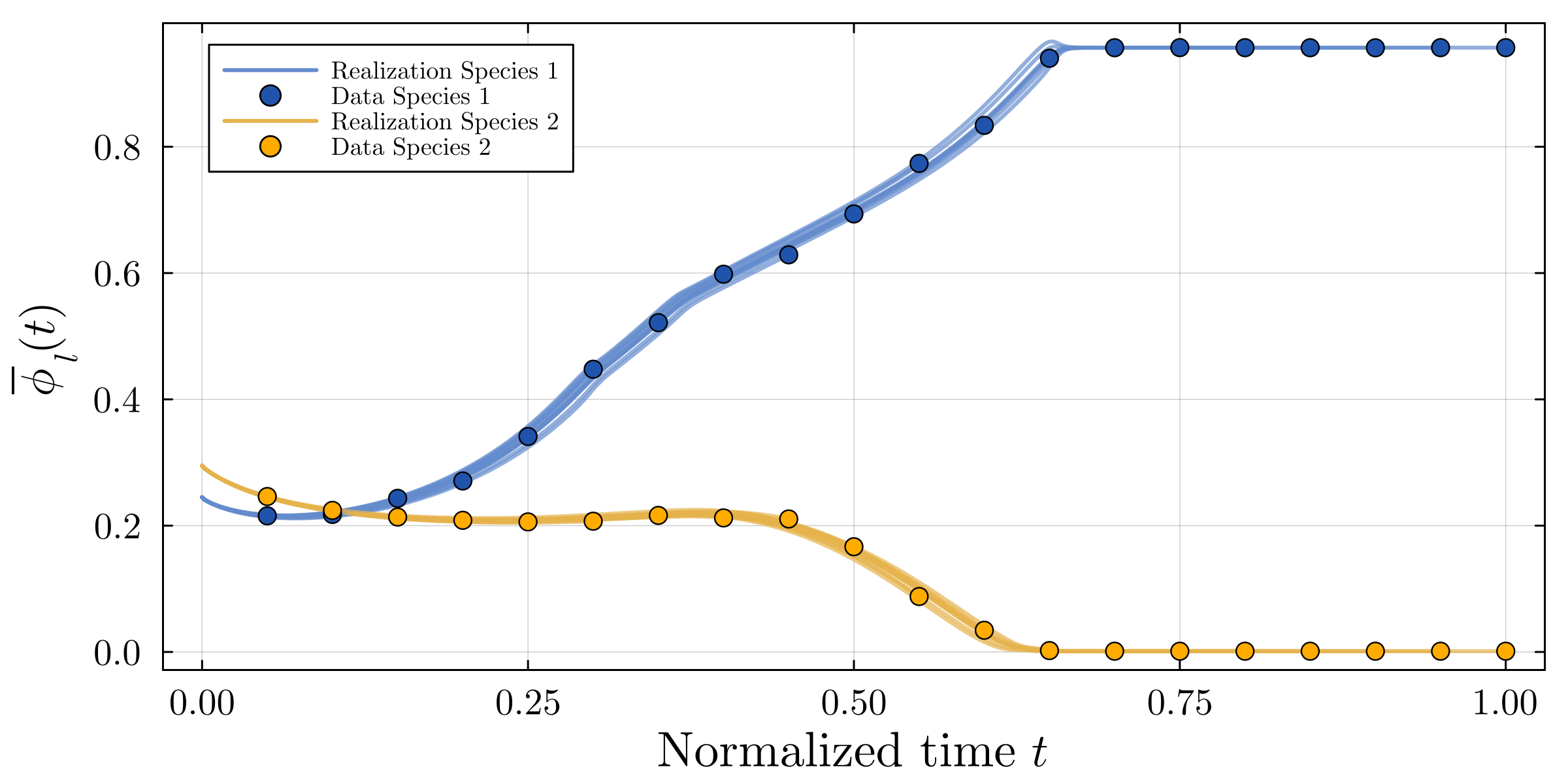}
    \caption{Dataset with $N_{\mathrm{data}} = 20$ volume fractions of living bacteria for two species, $\overline{\phi}_1(t)$ and $\overline{\phi}_2(t)$. Individual realizations are shown as yellow and blue lines, each ending at different steps, indicated by the dots. The data is generated from 1000 samples of the underlying true distribution of parameters with mean values $\ftheta^{*} = [1,\ 0.1,\ 1,\ 1,\ 2]$ along with \changed{$\mathrm{CoV}^a = 0.5 \%$}.}
    \label{fig:data-I}
\end{figure}

The dataset $\data$ comprises $N_{\mathrm{data}} = 20$ individual simulations, \changed{generated using mean values $\ftheta^{*} = [1,\ 0.1,\ 1,\ 1,\ 2]$ and the respective coefficients of variations, i.e., $\mathrm{CoV}^a$ and $\mathrm{CoV}^b$. Each simulation terminates at a different time step, as shown in \cref{fig:data-I}. Thus, the data is inherently uncertain since random samples are selected from different simulations with the same mean value $\ftheta^*$. In the updating process, the true mean value $\ftheta^*$ is assumed to be unknown. The goal is then to infer it from data.} 

The solid lines represent the time evolution of the volume fractions of living bacteria for both species, $\bar{\phi}_1$ and $\bar{\phi}_2$. The scatters present the randomly selected realizations at $m = 20$ evenly spaced time steps within the interval $t \in [0, \ 1]$. Since the initial conditions at $t = 0$ are known, the first time step is chosen at $t = 0.05$, as depicted in \cref{fig:data-I}.
It is important to highlight that the dataset $\data$ only consist of the 20 discrete data points. 
Further, we note that the time steps were chosen as a balance between accuracy by using a large number of time instances to compare the data and realism by using as little data as possible.

The different constant simulation parameters used in case I, e.g., the \changedIKM{rate sensitivity} $\bm{\eta}$, different initial conditions, nutrients and antibiotics, are summarized in \cref{tab:fixed-case-I}.
Further, for all the following cases, we use a penalty term $K_{p} = 10^{-4} \frac{\mathrm{J}}{\mathrm{m}^3}$ and initial $\psi_i = 0.999$ for all species.

\begin{table}[t]
\centering
\caption{Values of simulation parameters for case I.}\label{tab:fixed-case-I}
\begin{tabular}{rllr}
    \toprule
     & Variable & Unit & Value \\
    \midrule
    \changedIKM{rate sensitivity} & $\eta_1$ & $[\frac{\mathrm{kg}}{\mathrm{ms}}]$ & 1 \\
    \changedIKM{rate sensitivity} & $\eta_2$ & $[\frac{\mathrm{kg}}{\mathrm{ms}}]$ & 2 \\
    \midrule
    initial & $\phi_1$ & [-] & 0.25 \\
    initial & $\phi_2$ & [-] & 0.30 \\
    \midrule
    nutrients & $c^*$ & $[\frac{\mathrm{m}^2}{\mathrm{s}^2}]$ & 100\\
    antibiotics & $\alpha^{\star}$ & $[\frac{\mathrm{m}^2}{\mathrm{s}^2}]$ & 10 \\
    \midrule
    number of time steps & $N$ & [-] & $1000$  \\
    time step size & $\Delta t$ & $[\mathrm{s}]$ & $10^{-4}$ \\
    \midrule
    number of data points & $N_{\mathrm{data}}$ & [-] & $20$ \\
    number of aleatory samples & $N_\mathrm{samples}$ & [-] & $500$\\
    number of posterior samples & $N_\mathrm{posterior}$ & [-] & $5000$\\
    \midrule
    \changed{coefficient of variation $a$} & $\mathrm{CoV}^{a}$ & $[\%]$ & $0.5$\\
    \changed{coefficient of variation $b$} & $\mathrm{CoV}^{b}$ & $[\%]$ & $2.0$\\
    \botrule
\end{tabular}
\end{table}

\begin{table}[t]
\centering
\caption{Uniform prior ranges $\mathcal{U}(a, b)$ for the mean values of the parameters in case I.}\label{tab:priors-case-I}
\begin{tabular}{l c c c c c}
\toprule
Parameter & $\theta_1 = a_{11}$  & $\theta_2 = a_{12}$ & $\theta_3 = a_{22}$ & $\theta_4 = b_{1}$ & $\theta_5 = b_{2}$ \\
\midrule
Range  & $(0, 3)$  & $(0, 0.5)$ & $(0, 3)$ & $(0, 3)$ & $(0, 3)$  \\
\bottomrule
\end{tabular}
\end{table}

\begin{figure}[t]
    \centering
    \includegraphics[width=\linewidth]{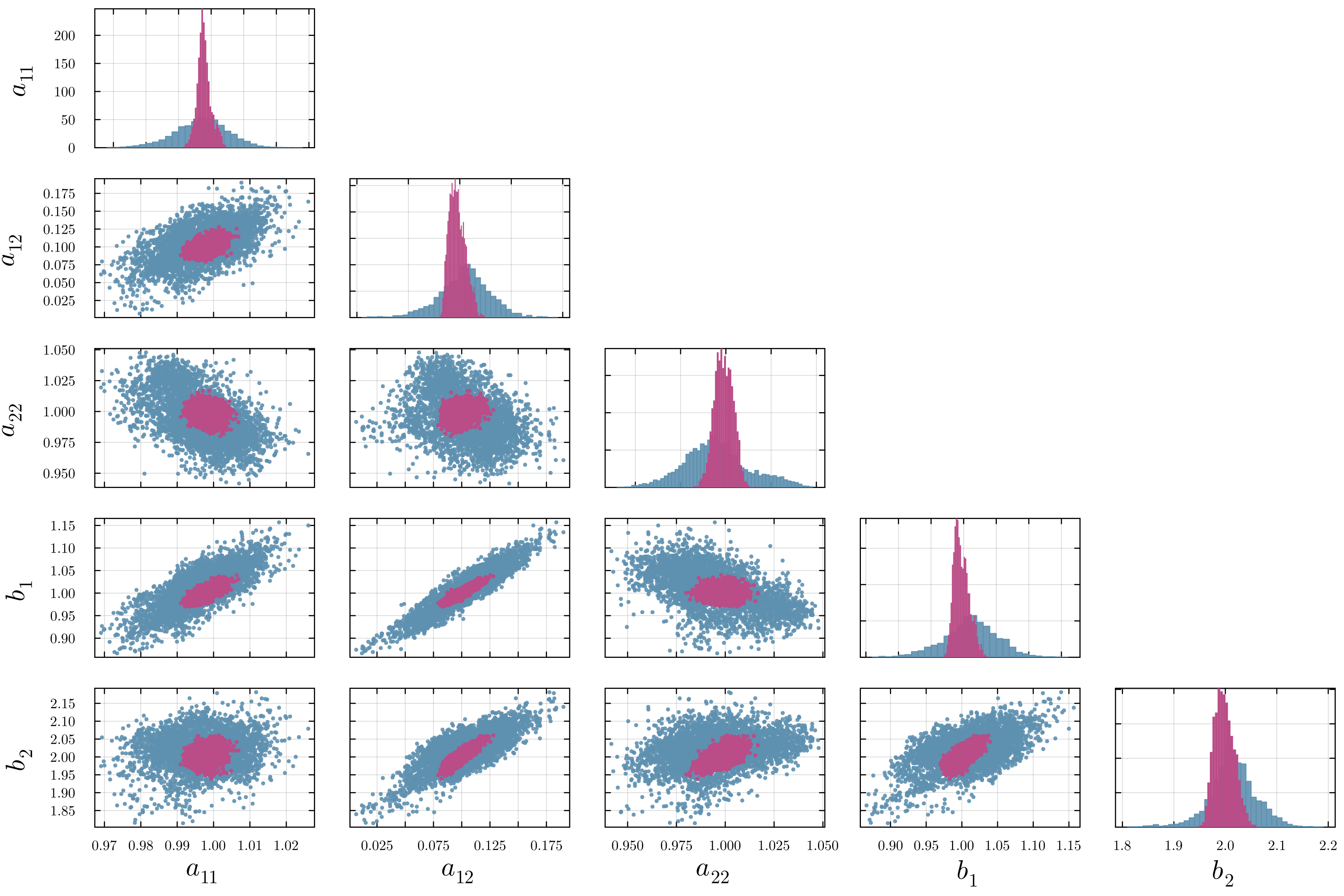}
    \caption{Comparison of posterior samples of the mean values of the five material parameters $\ftheta = [a_{11}, \ a_{12}, \  a_{22}, \  b_{1}, \  b_{2}]$ of case I calibrated with a model with $\mathrm{CoV}^a = 0.5 \%$ (purple) and $\mathrm{CoV}^b = 2 \%$ (blue).}
    \label{fig:post-samples-I}
\end{figure}

The Bayesian model updating with the described procedure is performed to calibrate the \changed{mean values of the parametric p-boxes describing the} model parameters.
The resulting samples of the posterior are visualized in \cref{fig:post-samples-I}, \changed{comparing the samples from both coefficients of variations and using the likelihood formulation proposed in \cref{eq:mvLogLkl_Biofilm}.} We observe that the posterior bounds \changed{for both CoVs} are much tighter compared to the priors, indicating a significant reduction in \changed{epistemic} uncertainty due to the incorporation of data. The posterior distributions generally peak around the ``true" parameter \changed{mean values $\ftheta^*$} used to construct the dataset, while still capturing some spread that reflects the inherent variability in the data. \changed{The results demonstrate that our Bayesian updating framework is capable of accurately resolving these values.
Further, it can be observed that in the case of the smaller $\mathrm{CoV}^a$, the identified posterior values are more sharply peaked than for the larger $\mathrm{CoV}^b$. This is to be expected due to the larger aleatory uncertainty involved in both the model predictions and the data set used for the model calibration. We see that the parameters can be inferred robustly when the aleatory uncertainty is increased.}

Moreover, dependencies between parameters become apparent, both in the scatter plots and in the Pearson correlation coefficients $\rho$ calculated from the posterior samples \changed{obtained with from the setup with $\mathrm{CoV}^a$}. For instance, we observe strong linear correlations between $a_{12}$ and $b_1$ ($\rho = 0.934$), as well as between $a_{12}$ and $b_2$ ($\rho = 0.898$). Moderate correlations are also evident between $a_{11}$ and $b_1$ ($\rho = 0.725$), and between $b_1$ and $b_2$ ($\rho = 0.759$).  \changed{The same correlations are observed for both the setup with small and the setup with large aleatory uncertainty.}
When considering the governing equations of the biofilm growth given by \cref{eq:EvolEq1,eq:EvolEq2,eq:EvolLagrange}, these correlations seem reasonable as the respective parameters jointly influence the growth and degradation behavior of the biofilm.

\begin{figure}[tp]
    \centering
    \begin{subfigure}{0.8\textwidth}
        \includegraphics[width=\linewidth]{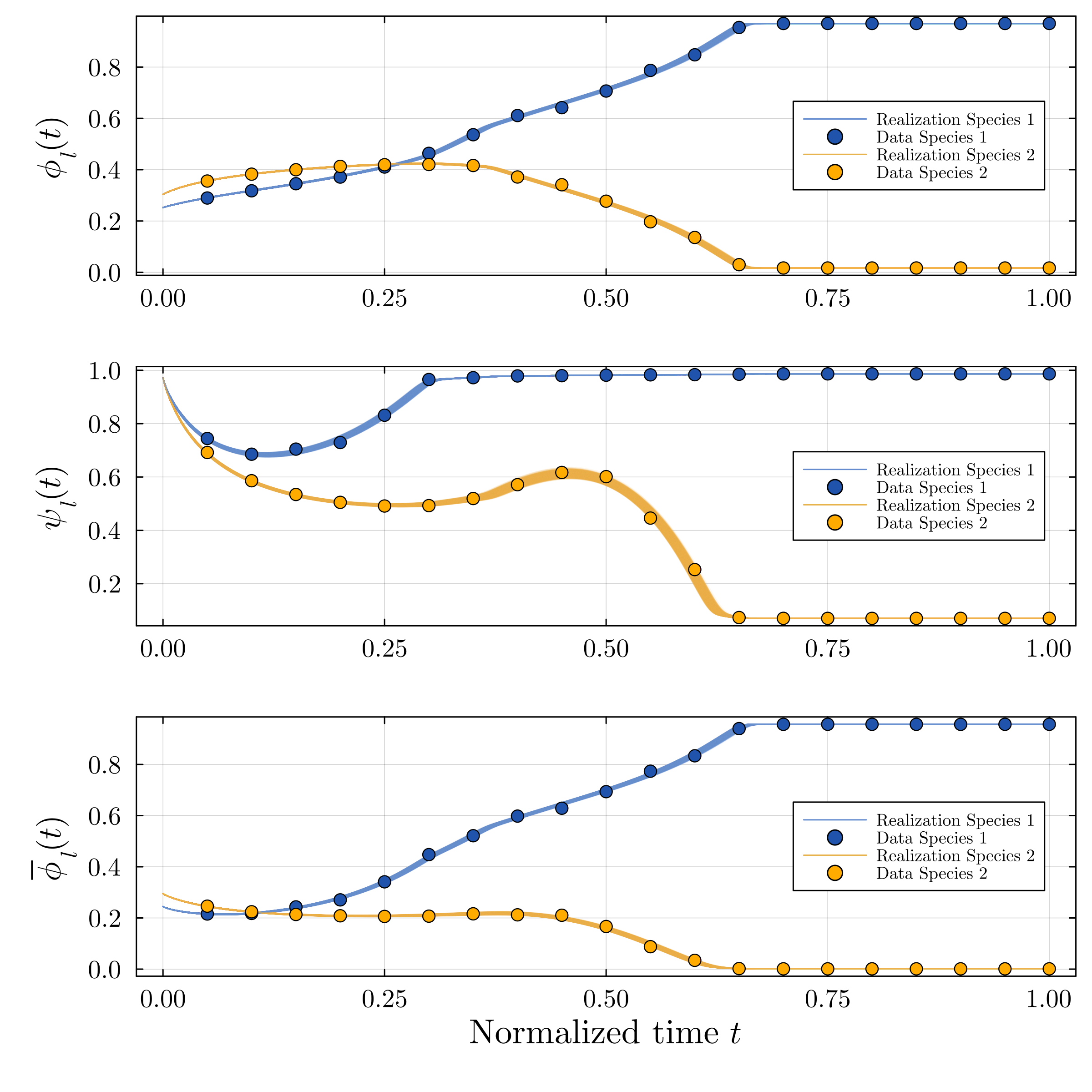}
        \caption{Posterior output corresponding to $\mathrm{CoV}^a$}
        \label{fig:post-output-I}
    \end{subfigure}
    \begin{subfigure}{0.8\textwidth}
        \includegraphics[width=\linewidth]{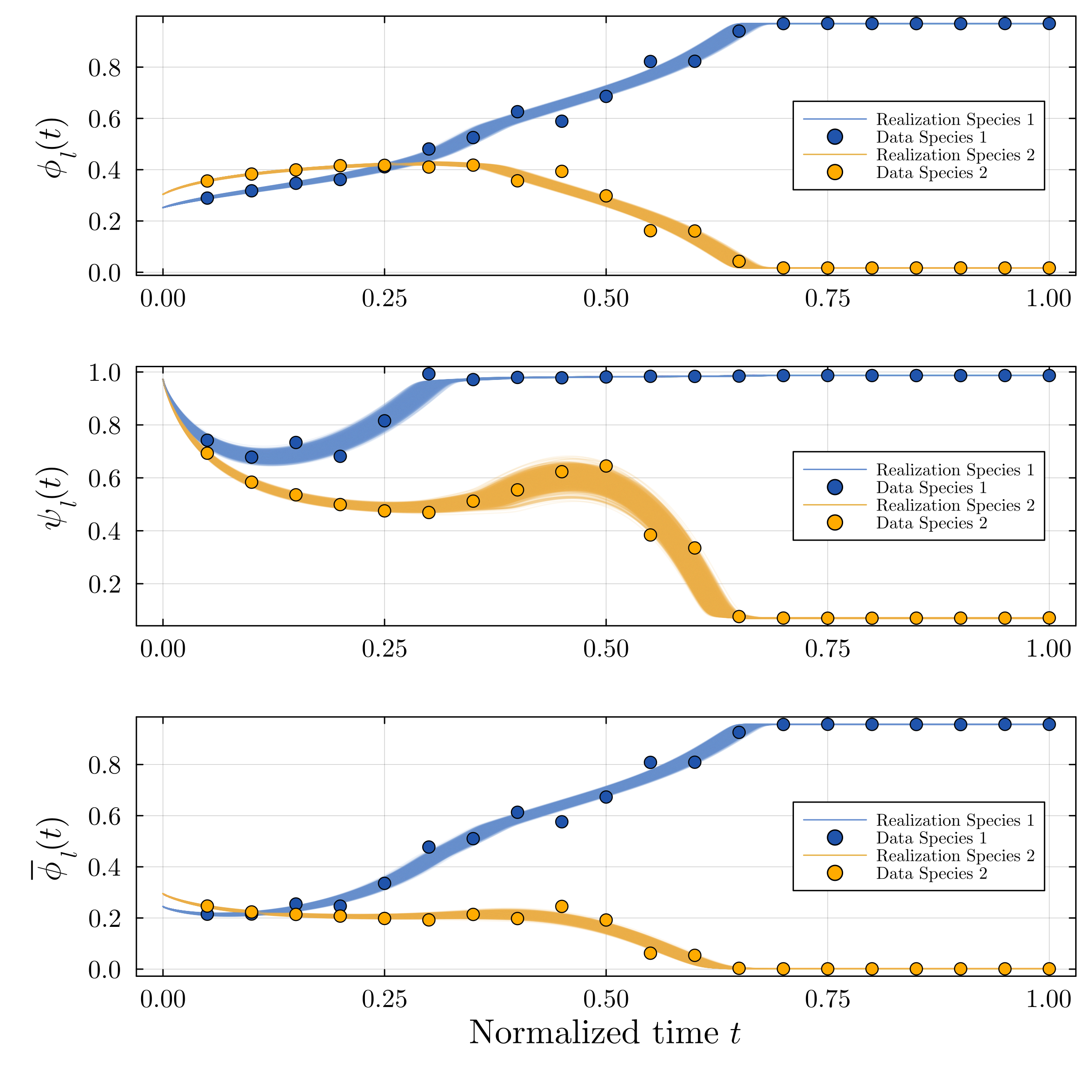}
        \caption{Posterior output corresponding to $\mathrm{CoV}^b$}
         \label{fig:post-output-I-higher}
    \end{subfigure}
    \caption{\changed{Comparison of the posterior output obtained with $\mathrm{CoV}^a$ (\cref{fig:post-output-I}) and $\mathrm{CoV}^b$ (\cref{fig:post-output-I-higher}).}}
    \label{fig:post-output-comparison}
\end{figure}

\Cref{fig:post-output-I} visualizes the model responses corresponding to the posterior samples, alongside the data $\data$ used for the updating \changed{and compares the results obtained from the two different setups. The figure visualizes the response corresponding to the 5000 posterior mean samples}.
There, we show the responses for the two individual quantities $\bm{\phi}$ and $\bm{\psi}$ as well as the combined measure $\Bar{\bm{\phi}} = \bm{\phi} \bm{\psi}$.
It shall be noted that the updating was only performed using the output $\Bar{\bm{\phi}}$ and the corresponding data. We visualize $\bm{\phi}$ and $\bm{\psi}$ along with the respective data to validate the calibrated model.
In general, we observe \changed{an excellent} agreement between the model outputs and the data: the posterior-informed simulations can reproduce similar behavior as the observed dataset while appropriately handling the variability introduced by random realizations and differing termination times across trajectories.
This holds true for all measures, also for the two quantities used for the validations which highlights the robustness of the parameter estimation.

\begin{figure}[t]
    \centering
    \includegraphics[width=0.8\linewidth]{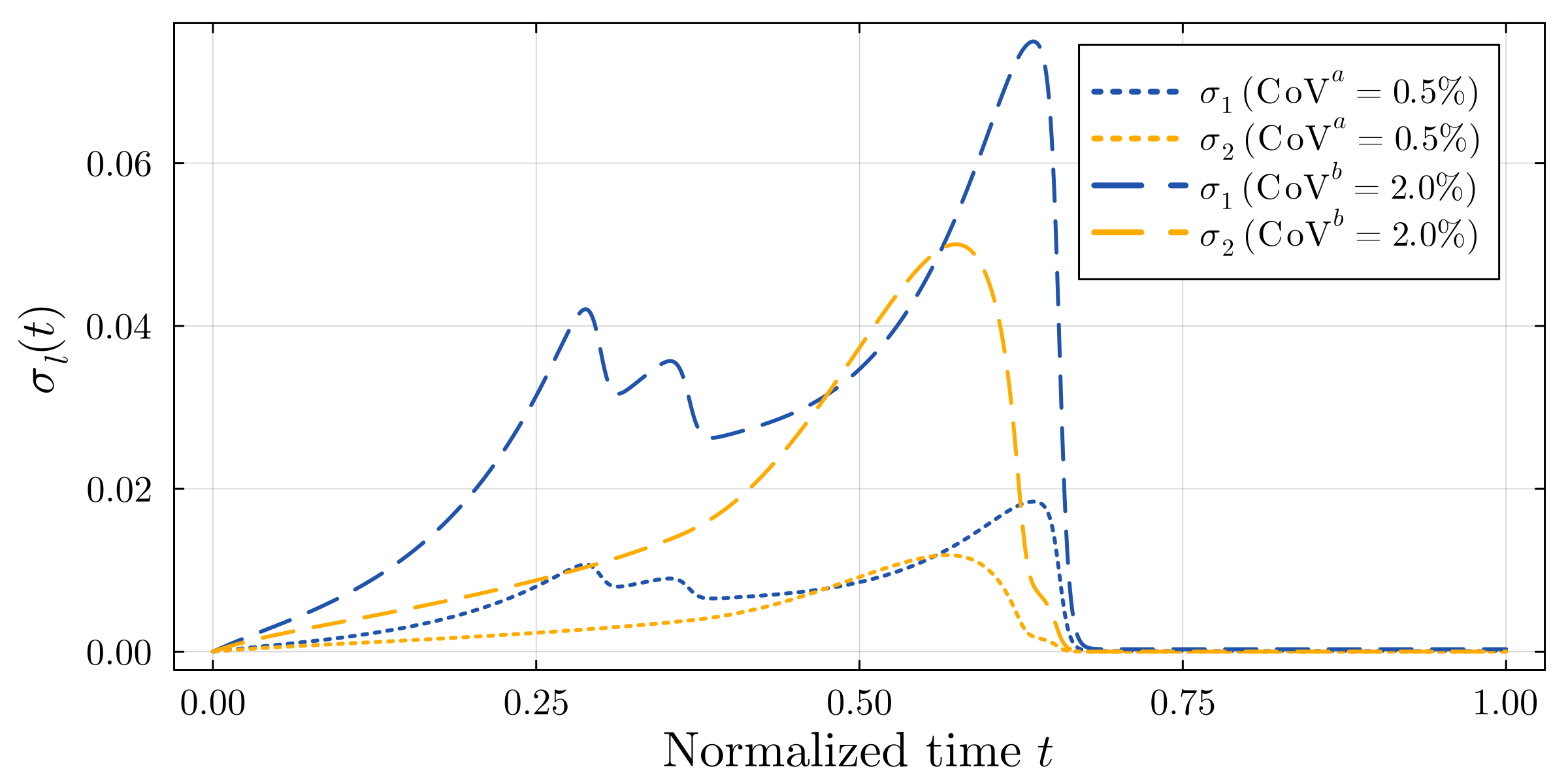}
    \caption{\changed{Comparison of the standard deviation of the outputs $\Bar{\phi}_l$ over time, $\sigma_l(t)$, for both coefficients of variation.}}
    \label{fig:sigma-comparison}
\end{figure}

\changed{
Evidently, the output variation is much higher, when the input variation is increased.
Furthermore, we can observe that the variation of the output is much higher than the variation in the input parameters which is likely attributed to the non-linearity of the model.
To show this, we compare the standard deviation of the response variable $\Bar{\phi}_l$ for the two species obtained for the different input coefficients of variation in \cref{fig:sigma-comparison}. The standard deviation is obtained from the TSM solution with $N_{\mathrm{samples}} = 1000$ samples and the true parameter mean $\ftheta^*$.
As also clear from the output visualization in \cref{fig:post-output-I}, the standard deviations varies with time.
When relating the standard deviation to the mean value, we can also calculate a coefficient of variation of the response.
For the setup with input variation $\mathrm{CoV}^a = 0.5 \% $, a maximum variation of the output quantity $\Bar{\phi}_1$ $\mathrm{CoV} = 2.61\% $ is observed. For $\Bar{\phi}_2$, we observe $\mathrm{CoV} = 54.40\%$ which can be attributed to $\Bar{\phi}_2$ converging to a smaller (mean) value at around $t \geq 0.6$.
For the input  $\mathrm{CoV}^b = 2 \%$, a we obtain maximum output variations of $\mathrm{CoV} = 10.48\%$ and $\mathrm{CoV} = 167.01\%$ for the measured $\Bar{\phi}_1$ and $\Bar{\phi}_2$, respectively.
}

The tight output prediction interval observed in \cref{fig:post-output-I} is a contrast to \cref{fig:prior-output-I}. \changed{As opposed to the wide range of solutions covered by the prior, the posterior output matches the data closely.}

\begin{figure}[t]
    \centering
    \begin{subfigure}{0.49\textwidth}
        \includegraphics[width=1\linewidth]{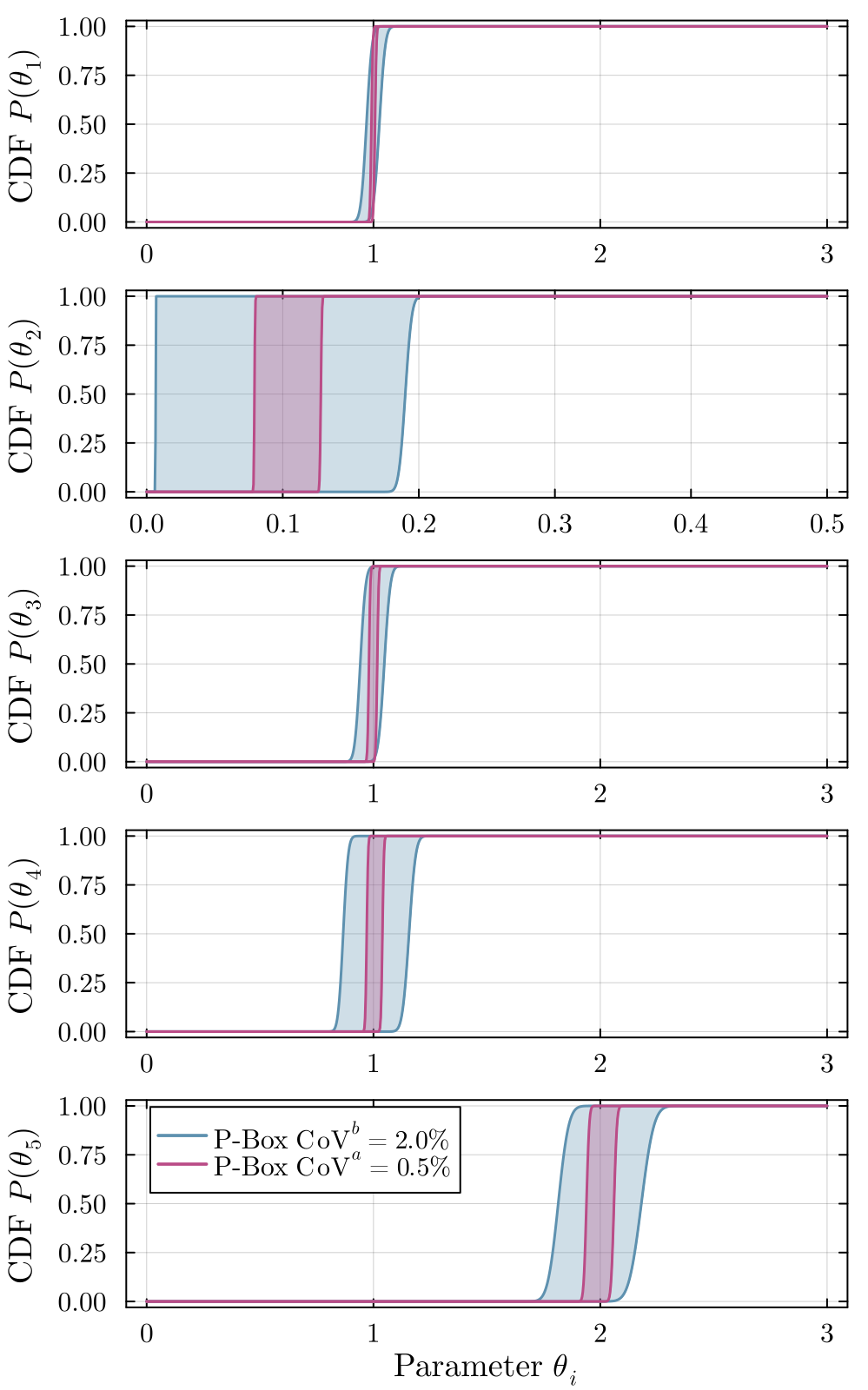}
        \caption{Diagonal covariance}
        \label{fig:pboxes_diagonalCovariance}
    \end{subfigure}
    \begin{subfigure}{0.49\textwidth}
        \includegraphics[width=1\linewidth]{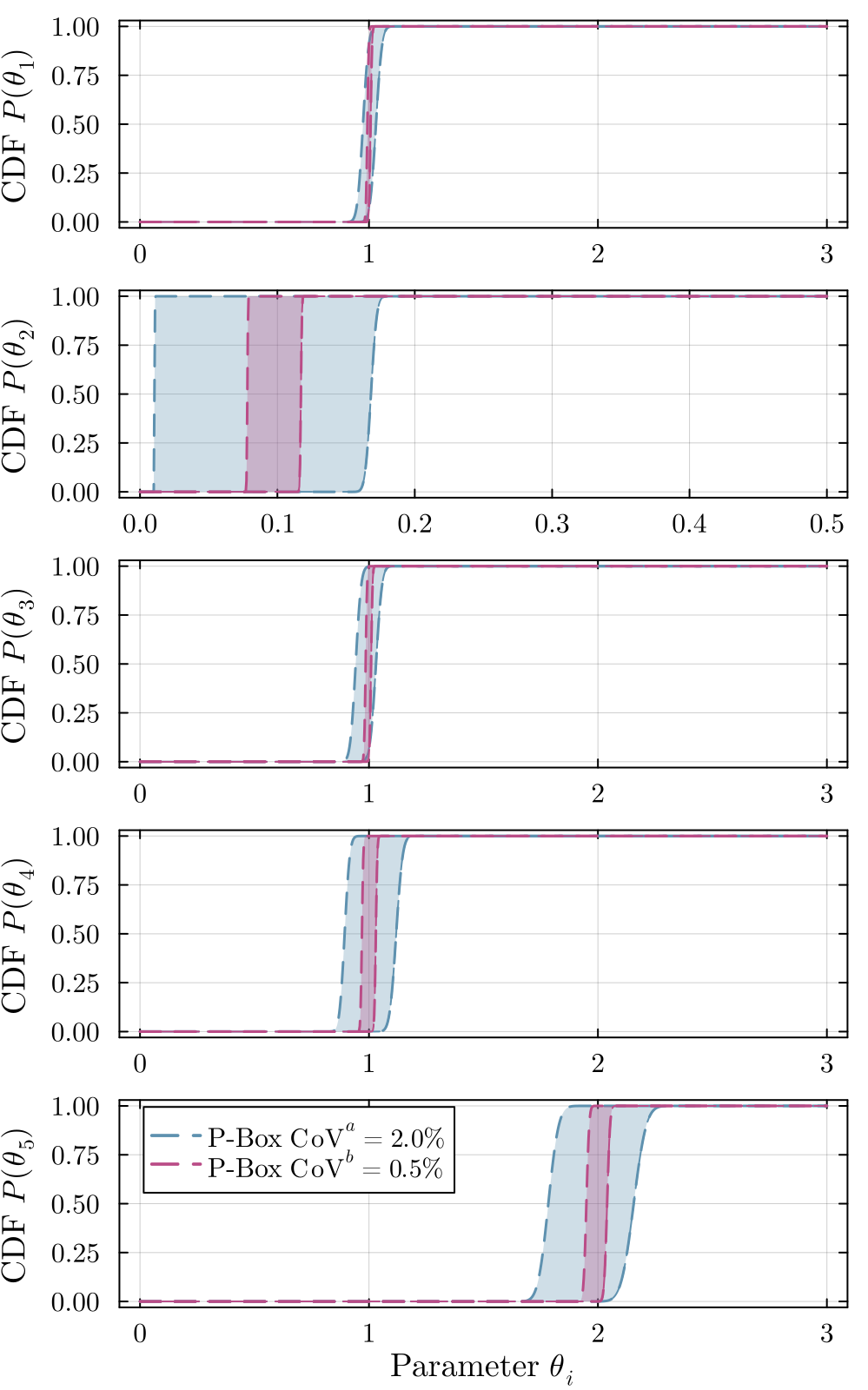}
        \caption{Full covariance}
        \label{fig:pboxes_fullCovariance}
    \end{subfigure}
    \caption{\changed{Comparison of the} updated p-boxes between using the diagonal covariance (\cref{fig:pboxes_diagonalCovariance}) and the full covariance (\cref{fig:pboxes_fullCovariance}) in the likelihood \changed{obtained with $\mathrm{CoV}^a$ (blue) and $\mathrm{CoV}^b$ (purple).}}
    \label{fig:pboxes}
\end{figure}

Lastly, to show that using only the diagonal elements of the estimated covariance matrix does not decrease the accuracy of the updating, \cref{fig:pboxes} shows the resulting p-boxes.
\changed{The figure compares the p-boxes obtained after the updating with the two different levels of uncertainty.}
\Cref{fig:pboxes_diagonalCovariance} shows results with only the diagonal elements, while \cref{fig:pboxes_fullCovariance} shows results obtained from using the full covariance. Both results do not diverge from each other, and the resulting intervals are very similar. Thus, in the following, we will concentrate on only using the diagonal elements.
\changed{Generally, the p-box obtained from updating with the smaller aleatory uncertainty is enclosed by the p-box obtained with larger uncertainty.}



\subsubsection{Computational efficiency and accuracy of the approach}
\label{sec:CompEff}
The model and its updating procedure are implemented in the open-source programming language \emph{Julia}.
Performance tests for a single deterministic model prediction and for the TSM-ROM implementation indicate that one model run with two species requires approximately \SI{3}{ms}, whereas solving the complete TSM-ROM takes about \SI{40}{ms} on the same hardware. 
\changedIKM{In total, 6 computations over all time steps are required for the TSM-ROM. One simulation for the zeroth order term in Equation~\eqref{eq:ZerothOrder} and five simulations for all first order terms, see Equation~\eqref{eq:FirstOrder}. After construction, the TSM-ROM allows to evaluate biofilm model for different parameter values in negligble computation time (in the order of a few \si{\micro \second}).
In comparison, MC simulations typically require often hundreds of individual model runs for convergent results. In the same time as the TSM-ROM takes, only $\approx 13$ individual simulations could be finished. These are not enough for convergent results.
To quantify the accuracy of the TSM-ROM, we compare the results of the TSM-ROM with evaluations of the true equation system for the two species model for 1000 different parameter values.
We compute the absolute error for each time step as
\begin{equation}
    \epsilon_\textrm{TSM-ROM, Time}(k) = \sum_{i=1}^{N_{\mathrm{samples}}} \sqrt{\left( \bar{\Phi}_1^k - \bar{\Phi}_1^k \right)^2 + \left(\bar{\Phi}_1^k - \bar{\Phi}_1^k \right)^2}
\end{equation}
The total absolute error over all time results as
\begin{equation}
    \epsilon_\textrm{TSM-ROM} = \frac{1}{N} \sum_{i=1}^N \epsilon_\textrm{TSM-ROM, Time}(i).
\end{equation}
For a COV of 2.0\%, an absolute error of $\epsilon_\textrm{TSM-ROM} = \num{5.1E-3}$ results and for a COV of 0.5\%, $\epsilon_\textrm{TSM-ROM} = \num{2.2E-4}$. The total range of $\bar{\Phi}$ is inbetween 0 and 1.
The absolute error over time $\epsilon_\textrm{TSM-ROM, Time}$ is presented in Figure~\ref{fig:AbsError} for the case of a CoV of 2.0\%.
Typically, the largest error occurs when a biofilm a total volume of 1. Here, a high degree of nonlinearity occurs and a linear Taylor series seems not sufficient. Nevertheless, even at these time points, the absolute error is only $\textrm{max}(\epsilon_{\textrm{TSM-ROM, Time}}) =\num{5.4E-2}$ for a coefficient of variation of 2.0 \%.
} 

\changedIKM{
\begin{figure}
    \centering
    \includegraphics[width=0.8\linewidth]{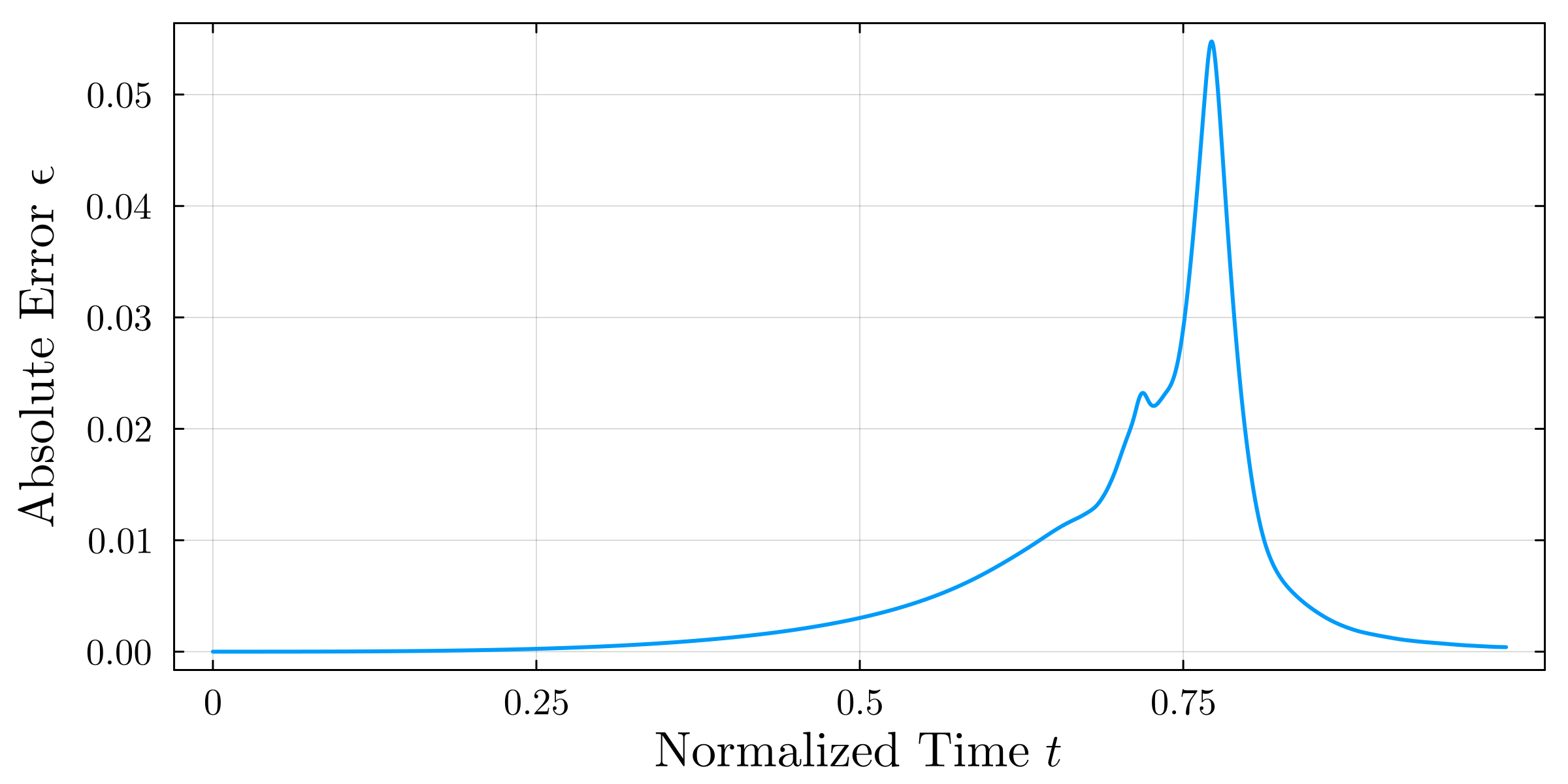}
    \caption{Absolute error $\epsilon_\textrm{TSM-ROM, Time}$ over time for the case of CoV of 2.0\%.}
    \label{fig:AbsError}
\end{figure}
}

\subsection{Case II: Four-Species Biofilm Model}
\begin{figure}[t]
    \centering
    \includegraphics[width=\linewidth]{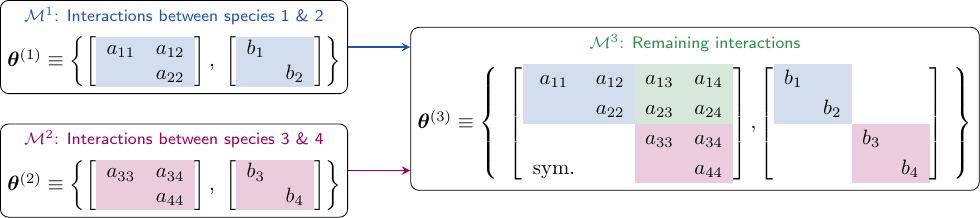}
    \caption{Visualization of the hierarchical updating procedure on two different levels: Updating of two model $\model^1$ and $\model^2$ is performed using simpler two-species models, subsequently $\model^3$ is used to update the remaining interactions.}
    \label{fig:hierarchical-update}
\end{figure}

\changed{As a second case study, we consider a biofilm model with $n=4$ interacting species and a total of 14 unknown parameters. To render Bayesian inference tractable, we employ a hierarchical (multilevel) updating strategy \cite{behmaneshHierarchicalBayesianModel2015} in three stages. Such a staggered approach updates model parameters sequentially rather than updating all fourteen parameters at once. This modified setup is employed to improve efficiency and convergence of the TMCMC algorithm which decreases dramatically as the dimensionally, i.e., the number of unknown parameters, increases. Further information is found in the study in Ref. \cite{betz2016}, where the authors highlight the impact of parameter dimension on the performance of TMCMC algorithms. The nonlinear nature of the model adds further motivation for this hierarchical approach. Nonlinear dynamics can lead to multiple qualitatively different classes of outcomes, even under the same prior assumptions. For instance, in the two-species example of \cref{fig:prior-output-I}, prior samples may produce scenarios in which either species ends up with the higher local concentration. To obtain meaningful likelihood evaluations, it is therefore crucial that the model outputs generated from prior samples lie at least reasonably close to the observed data. Sequentially updating parameters helps steer the prior in stages toward regions of the parameter space where the model response is compatible with the data, thereby supporting a more stable and reliable inference process.} In our approach, we first decompose the full four-species system into two simpler two-species submodels: $\model^1$ captures the interactions of species 1 and 2 (five parameters), and $\model^2$ captures the interactions of species 3 and 4 (five parameters). These submodels are calibrated \emph{in parallel} using uninformative priors \changed{$\theta_i \sim \mathcal{U}(0,3)$}, yielding posterior distributions on the reduced parameter spaces $\mathcal{D}_{\Theta}^{1}\subset\mathbb{R}^5$ and $\mathcal{D}_{\Theta}^{2}\subset\mathbb{R}^5$. A schematic of this multilevel updating is shown in \cref{fig:hierarchical-update}. \changed{Note that the order of the species is unimportant in this example. It would also be possible to define $\mathcal{M}^1$ and $\mathcal{M}^2$ with different parameter combinations than the ones given here. Further, a staggered approach with first a two-species model, then a three-species model and finally the full four-species model would be possible.}

Next, in model $\model^3$, we assemble the full four-species interaction matrices $\bm{A}, \bm{B} \in \mathbb{R}^{4\times4}$, where the ten parameters already inferred in the submodels are highlighted (blue for $\model^1$, red for $\model^2$). We now fix the corresponding ten parameters at their respective maximum-a-posteriori (MAP) estimates obtained from Steps 1 and 2. As a result, only the four remaining cross-block interaction parameters ${a_{13}, a_{14}, a_{23}, a_{24}}$ are treated as uncertain in $\model^3$ and assigned uninformative uniform priors \changed{$\mathcal{U}(0,3)$}, yielding a reduced inference subspace $\mathcal{D}_{\Theta}^{3} \subset \mathbb{R}^4$. A final Bayesian update is then performed solely over this four-dimensional subspace.

In practice, Steps 1 and 2 follow the same two-species Bayesian updating procedure described in Case I, starting from uninformative priors. In Step 3, instead of reusing full posterior distributions, we carry forward only the MAP estimates from $\model^1$ and $\model^2$, thereby focusing inference in $\model^3$ entirely on the remaining four interaction parameters. This hierarchical structure reduces the dimensionality of each inference step, from 14 parameters in total to two problems of dimension 5, followed by one of dimension 4.

Afterwards, we look at a modified setup of $\model^3$ to validate the calibrated parameters with new data in a different setup. For this, we apply the antibiotics only after $t=0.5$ to check if our calibrated are model parameters are robust to this change in the setup. We denote this as the model $\model^3_{\mathrm{val}}$.
The selected simulation parameters of the submodels are summarized in \cref{tab:fixed-case-II}.

Values in \cref{tab:fixed-case-II} vary between $\model^1$, $\model^2$ and $\model^3$ since the different species have different sensitivities to nutrients and antibiotics. In $\model^2$ the antibiotics were reduced to not have a zero-concentration of the bacteria, since this would lead to non-informative data. Moreover, $\model^2$ has a longer experimental duration (5000 instead of 2500 time steps) due to slower growth of the microfilms. We simulated $\model^1$ for a shorter duration because the concentrations became almost stationary and did not result in any further information gain. \changed{Further, we restrict this example to a $\mathrm{CoV}$ of $0.5 \%$ as the uncertainty bounds on the two species model for $2 \%$ are already quite large and due to the increased number of parameters in this example would be even larger.}

\begin{table}[t]
\centering
\caption{Values of simulation parameters for the submodels of case II.}\label{tab:fixed-case-II}
\begin{tabular}{rllrrrr}
    \toprule
     & Variable & Unit & $\model^1$ & $\model^2$ & $\model^3$ & $\model^3_{\mathrm{val}}$ \\
    \midrule
    \changedIKM{rate sensitivity} & $\eta_1$ & $[\frac{\mathrm{kg}}{\mathrm{ms}}]$ & 1.0 & - & 1.0 & 1.0\\
    \changedIKM{rate sensitivity} & $\eta_2$ & $[\frac{\mathrm{kg}}{\mathrm{ms}}]$ & 1.0 & - & 1.0 & 1.0\\
    \changedIKM{rate sensitivity} & $\eta_3$ & $[\frac{\mathrm{kg}}{\mathrm{ms}}]$ & - & 1.0 & 1.0 & 1.0\\
    \changedIKM{rate sensitivity} & $\eta_4$ & $[\frac{\mathrm{kg}}{\mathrm{ms}}]$ & - & 1.0 & 1.0 & 1.0\\
    \midrule
    initial & $\phi_1$ & [-] & 0.2 & - & 0.02 & 0.02 \\
    initial & $\phi_2$ & [-] & 0.2 & - & 0.02 & 0.02 \\
    initial & $\phi_3$ & [-] & -    & 0.2  & 0.02 & 0.02 \\
    initial & $\phi_4$ & [-] & -    & 0.2  & 0.02 & 0.02 \\
    \midrule
    nutrients & $c^*$ & $[\frac{\mathrm{m}^2}{\mathrm{s}^2}]$ & 100 & 100 & 25 & 25\\
    antibiotics & $\alpha^{\star}$ & $[\frac{\mathrm{m}^2}{\mathrm{s}^2}]$ & 100 & 10 & 0 & $50 \ \mathbb{I}[t > 500]$ \\
    \midrule
    number of time steps & $N$ & [-] & $2500$ & $5000$ & $750$ & $1500$\\
    time step size & $\Delta t$ & $[\mathrm{s}]$ & $10^{-5}$ & $10^{-5}$ & $10^{-4}$ & $10^{-4}$\\
    \midrule
    number of data points & $N_{\mathrm{data}}$ & [-] & $20$ & $20$ & $20$ & $20$\\
    number of aleatory samples & $N_\mathrm{samples}$ & [-] & $500$ & $500$ & $500$ & $500$\\
    number of posterior samples & $N_\mathrm{posterior}$ & [-] & $5000$ & $5000$ & $5000$ & -\\
    \midrule
    coefficient of variation & $\mathrm{CoV}$ & $[\%]$ & $0.5$ & $0.5$ & $0.5$ & $0.5$\\
    \botrule
\end{tabular}
\end{table}

\subsubsection{Interaction of Species 1 and 2}

\begin{figure}[t]
    \centering
    \includegraphics[width=\linewidth]{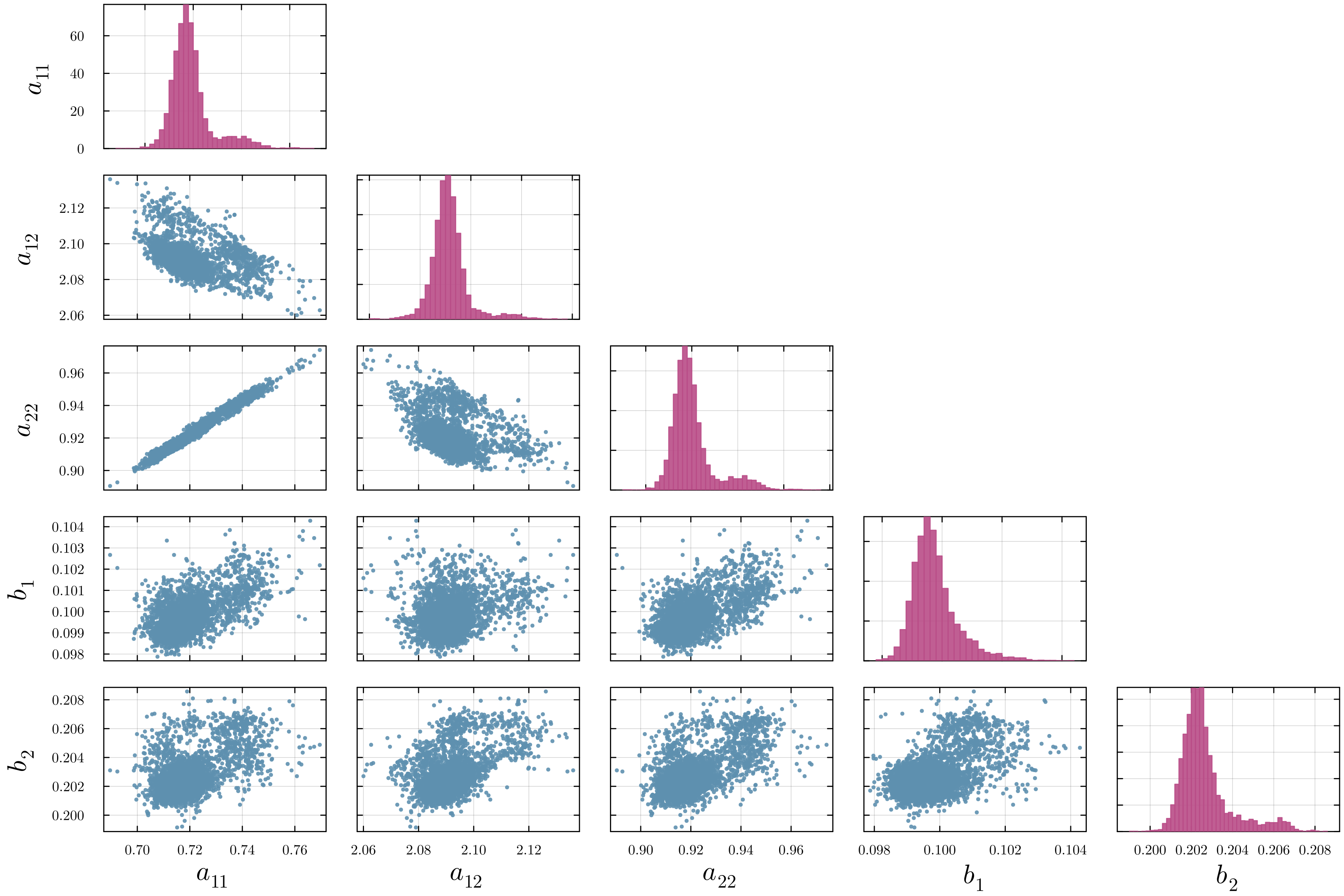}
    \caption{Posterior samples of the mean values of the five parameter in the set $\ftheta^{(1)}$ updated with the two-species model $\model^1$.}
    \label{fig:samples-M1}
\end{figure}

First, the parameter set $\ftheta^{(1)}$ is inferred using the two-species model $\model^1$. For this first submodel, the antibiotic parameter is set to $\alpha^{\changed{\star}} = 100 \ \frac{\mathrm{m}^2}{\mathrm{s}^2}$.
We employ the same likelihood and same approach as in case I for the updating.
The resulting posterior samples are shown in \cref{fig:samples-M1}. Here, again, for every parameter, a single peak along with some spread \changed{around} that can be observed. Notably, we observe a strong correlation between the mean values of the parameters $a_{11}$ and $a_{22}$.

\begin{figure}[t]
    \centering
    \includegraphics[width=\linewidth]{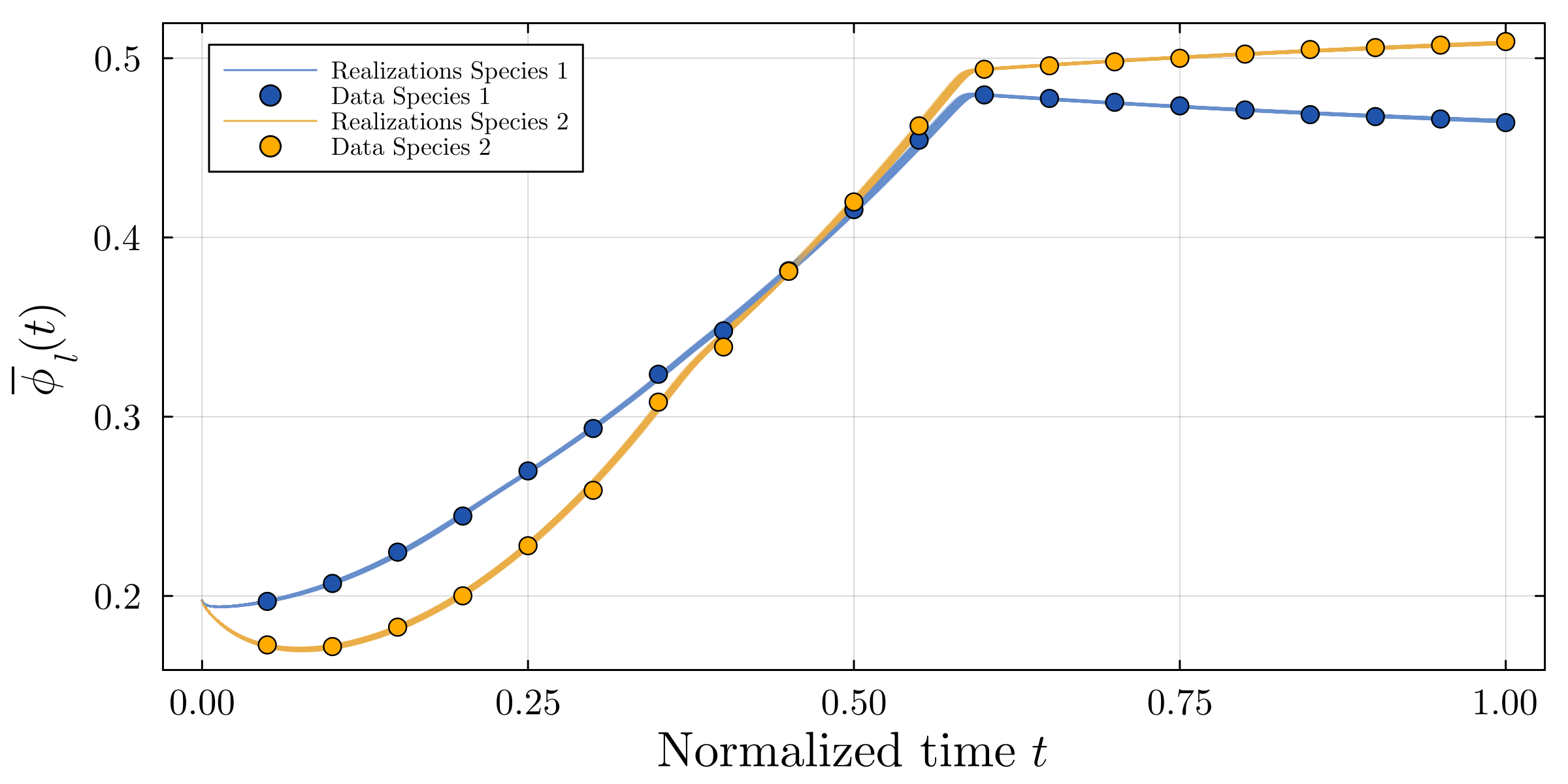}
    \caption{Comparison of the model output of model $\model^1$ corresponding to calibrated posterior samples (shaded) and the data (scatter).}
    \label{fig:output-M1}
\end{figure}

\Cref{fig:output-M1} visualizes the model output corresponding to the posterior samples along with the data points used for the updating. \changed{Excellent} agreement between the model response and the data can be observed.

\subsubsection{Interaction of Species 3 and 4}

\begin{figure}[t]
    \centering
    \includegraphics[width=\linewidth]{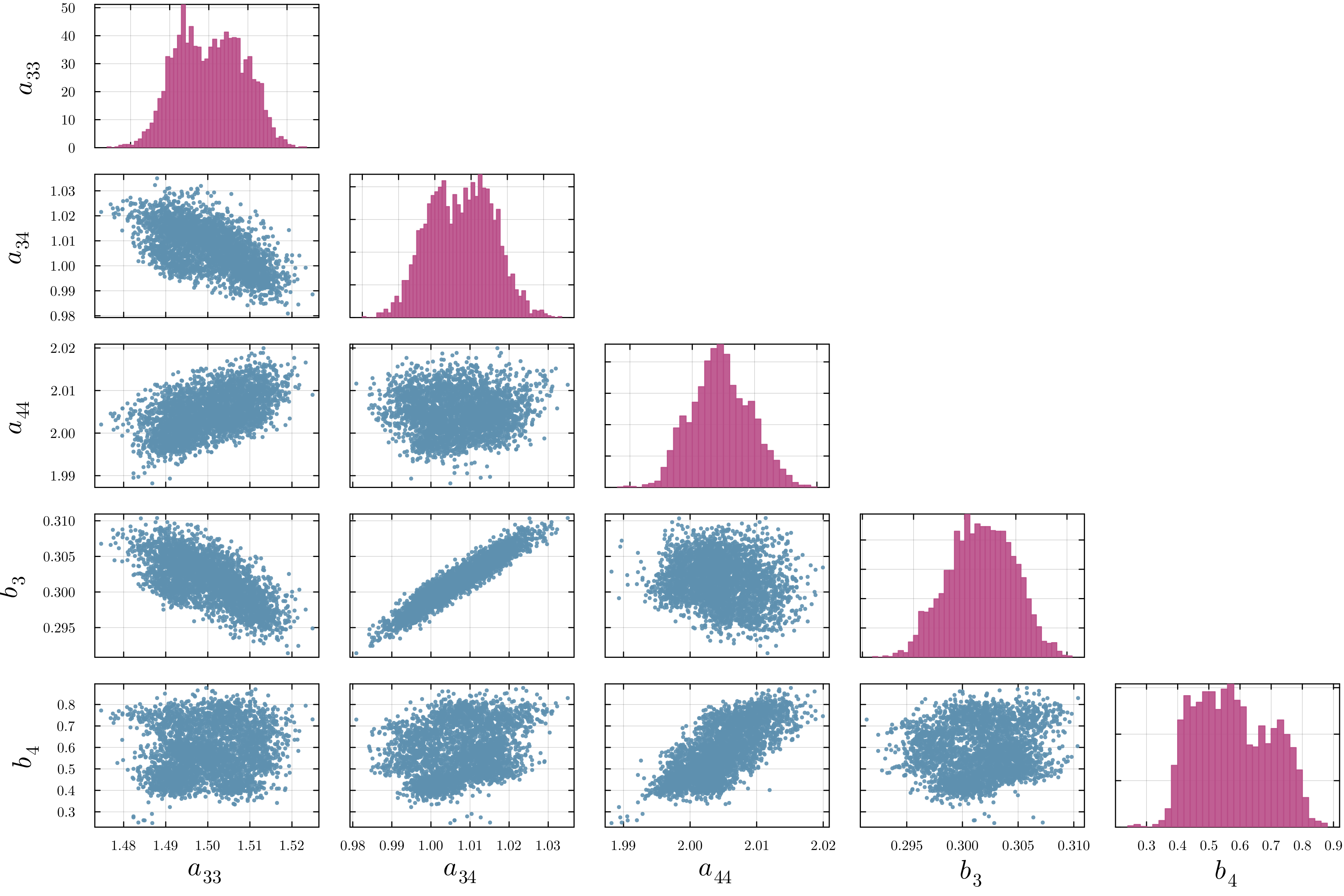}
    \caption{Posterior samples of the mean values of the five parameter in the set $\ftheta^{(2)}$ updated with the two-species model $\model^2$.}
    \label{fig:samples-M2}
\end{figure}

\begin{figure}[t]
    \centering
    \includegraphics[width=\linewidth]{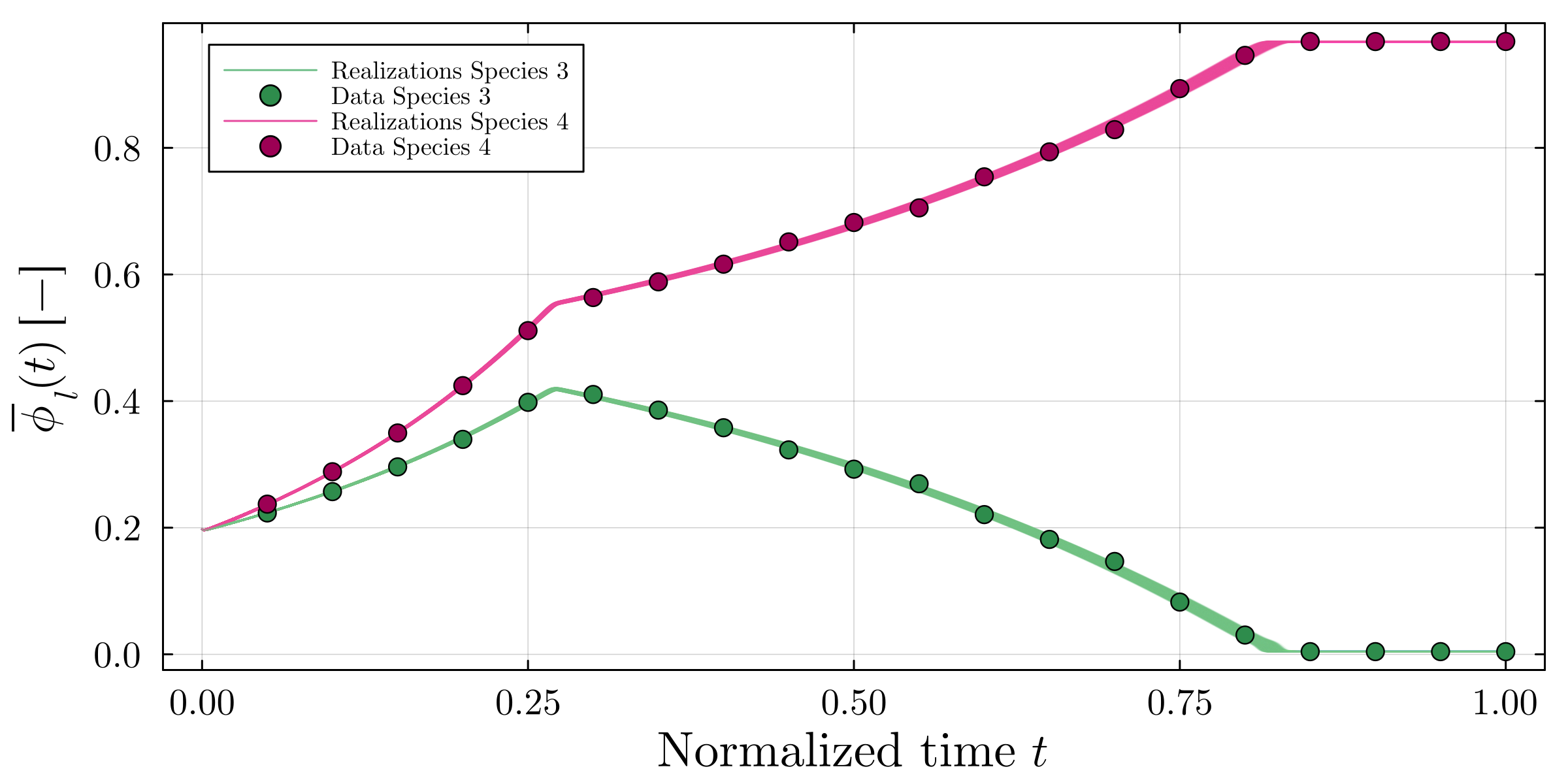}
    \caption{Comparison of the model output of model $\model^2$ corresponding to calibrated posterior samples (shaded) and the data (scatter).}
    \label{fig:output-M2}
\end{figure}

The same approach is applied to infer the mean values of the parameter set $\ftheta^{(2)}$ with the second two-species model, $\model^2$.
The results of the model calibration are shown in \cref{fig:samples-M2,fig:output-M2}, which again show the posterior samples and model outputs, respectively.
Here, an antibiotic concentration of $\alpha^{\changed{\star}} = 10 \ \frac{\mathrm{m}^2}{\mathrm{s}^2}$ is applied to build the data set and perform the updating.

It can be observed that the peak are not as sharp as in the case of the first parameter set. However, for all parameters but $b_4$, the spread around the peak is still small where comparing the ranges of the posterior samples to the prior range, i.e., $[0, \ 3]$. Only for the antibiotic sensitivity of the forth species, $b_4$, the range of posterior samples is rather large.

The model output in \cref{fig:output-M2}, however, shows \changed{excellent} agreement with the data points that are used for the Bayesian updating of the second parameter set.

\subsubsection{Remaining Interactions}

After the first two parameter sets are inferred using the two submodels $\model^1$ and $\model^2$, the remaining interaction parameters can be determined with the final four-species model $\model^3$.
In this setup, we fix the parameters in $\ftheta^{(1)}$ and $\ftheta^{(2)}$ to their respective MAP estimates.
Thus, only the remaining four interactions parameters are inferred, denotes as $\ftheta^{(3)}$. 
Since the remaining interaction parameters are only in the matrix $\bm{A}$, the term that is not dependent on the antibiotic concentration, we set the latter to $\alpha^{\changed{\star}} = 0$.

\begin{figure}[t]
    \centering
    \includegraphics[width=\linewidth]{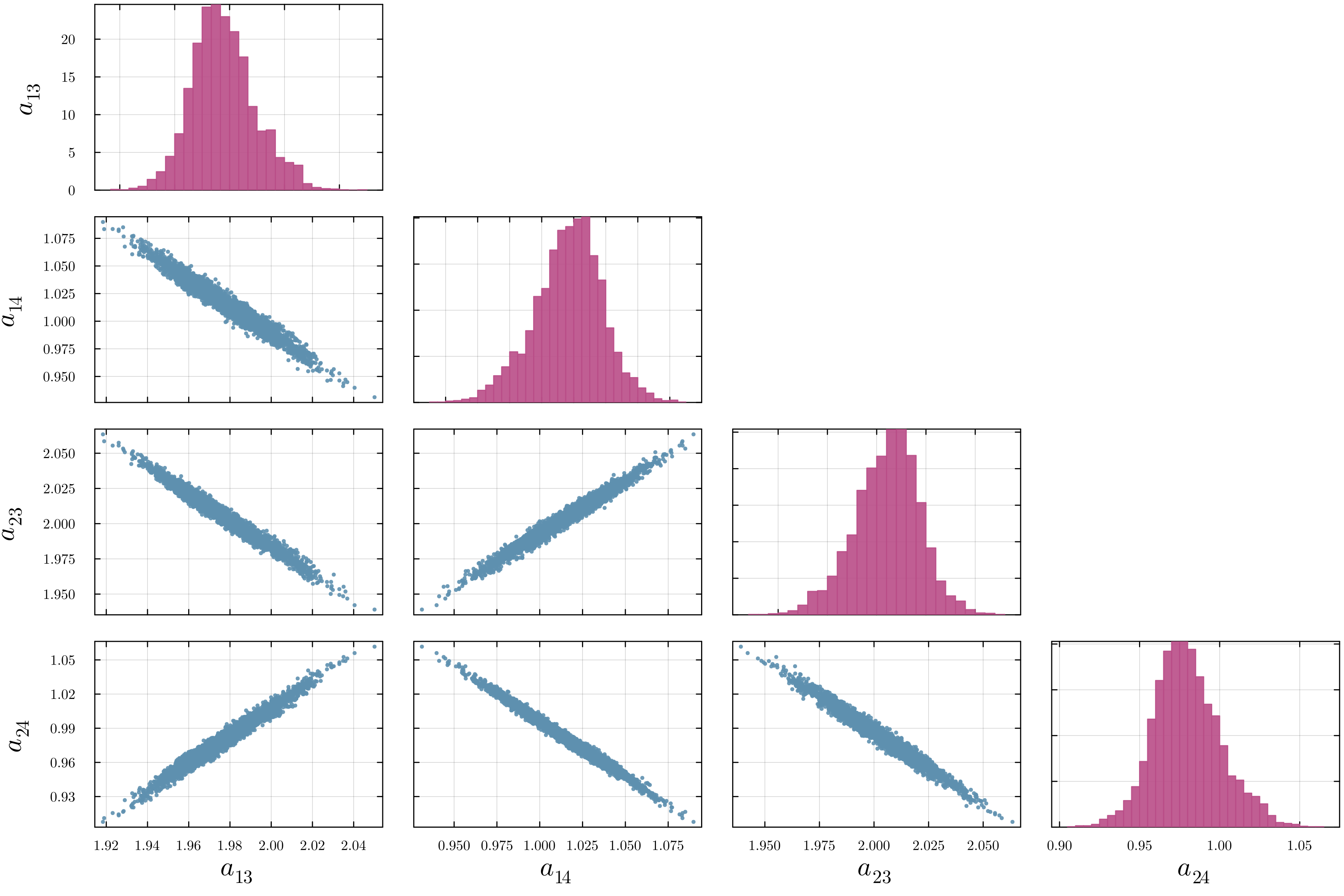}
    \caption{Posterior samples of the mean values of the parameter set $\ftheta^{(3)}$ updated with the four-species model $\model^3$.}
    \label{fig:samples-M3}
\end{figure}

\begin{figure}[t]
    \centering
    \includegraphics[width=\linewidth]{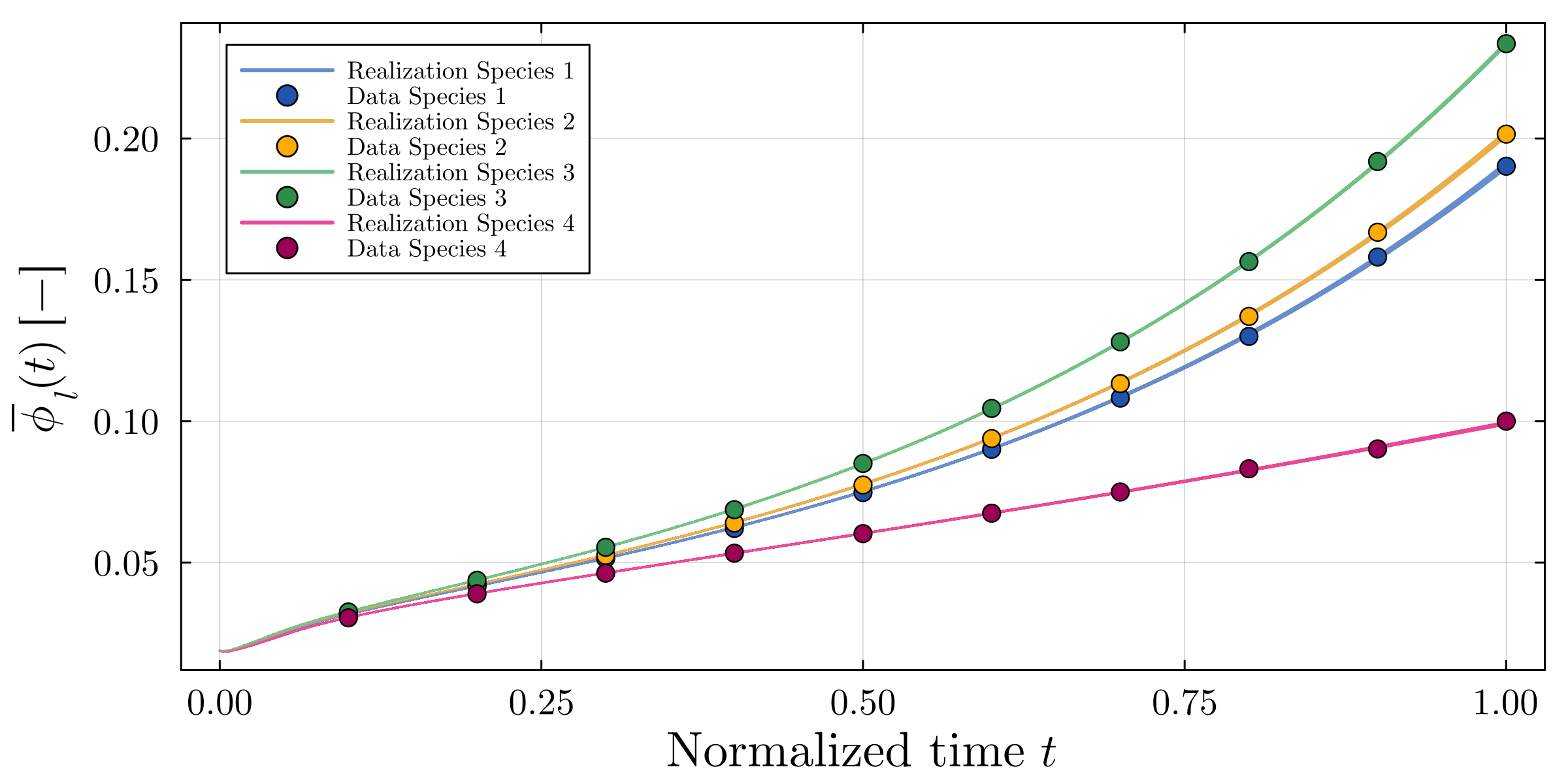}
    \caption{Comparison of the model output of model $\model^3$ corresponding to calibrated posterior samples (shaded) and the data (scatter).}
    \label{fig:output-M3}
\end{figure}

The results of this third and last updating are show in \cref{fig:samples-M3,fig:output-M3}.
Here, sharp and distinct peaks along with a strong linear correlation can be observed for all four parameters.
Again, a good agreement between posterior model response and data is observed after the updating.

\subsubsection{Comparison of the identified posterior mean with true mean}
\begin{figure}[tp]
    \centering
    \includegraphics[width=\linewidth]{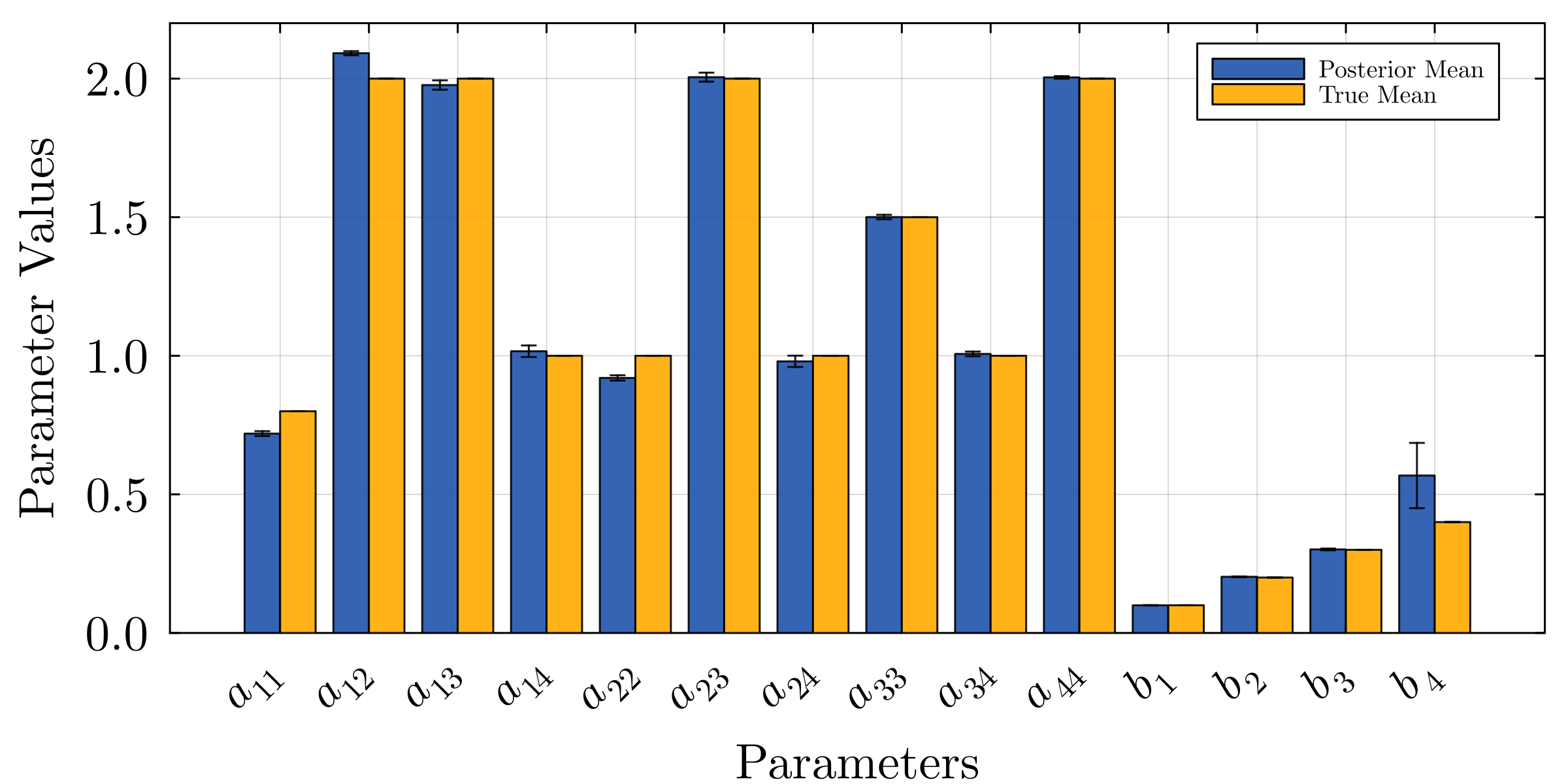}
    \caption{Comparison of the mean values of the identified parameters and true values used to generate the data used in case II. The error bars reflect the standard deviation of the posterior samples of the respective parameters.}
    \label{fig:posterior-comparison}
\end{figure}

In \cref{fig:posterior-comparison}, we compare the identified material parameters with the true parameter values used for the data generation. In addition, an error bar is added to the posterior means highlighting the standard deviation of the posterior. For almost all parameters, the values are very similar. The largest difference is obtained for the parameter $b_4$ connected to the sensitivity to antibiotics of the fourth biofilm. In addition, a rather high standard deviation of the posterior results. This gives an important hint that more data is needed for a better identification.

\begin{figure}[tp]
    \centering
    \includegraphics[width=\linewidth]{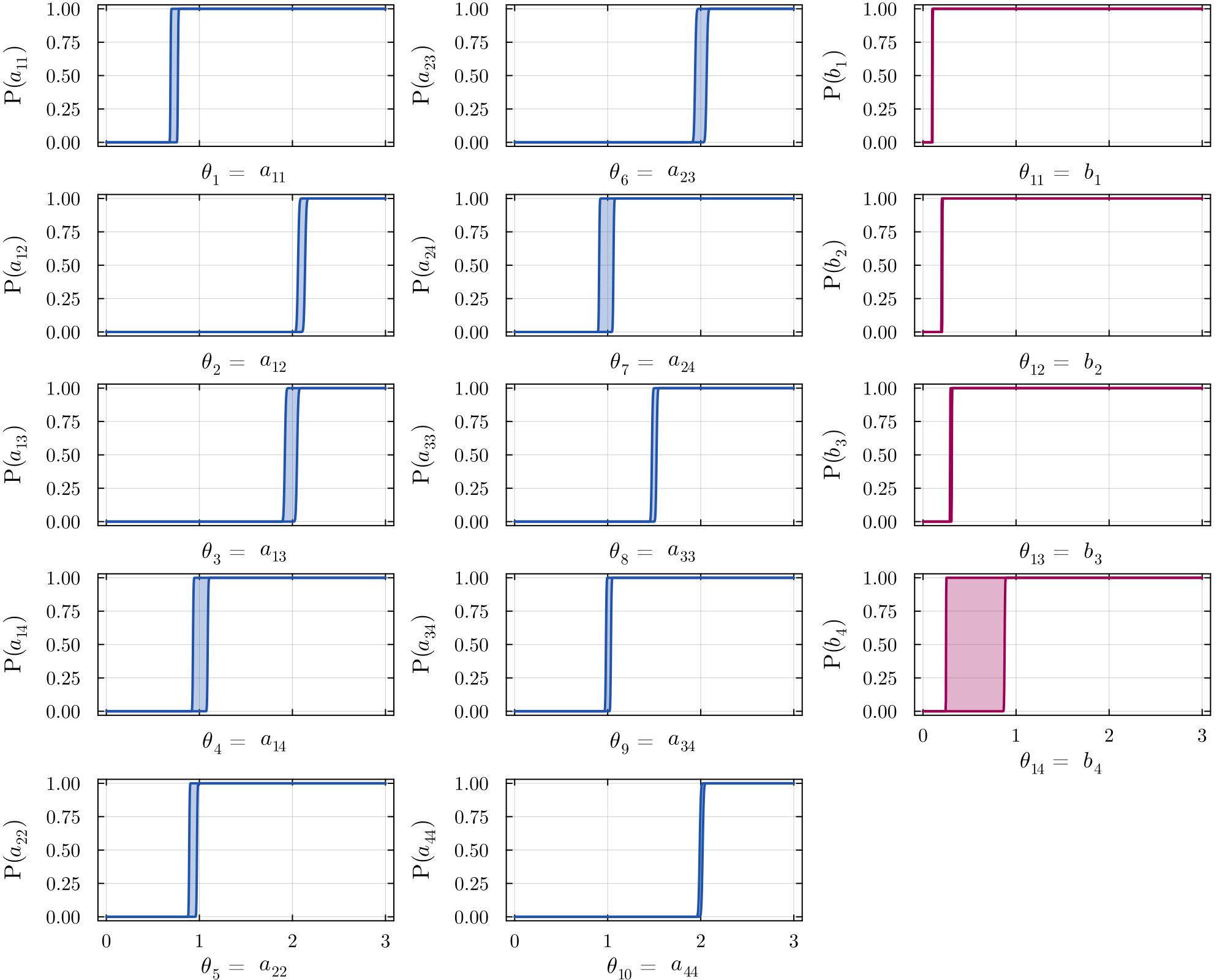}
    \caption{\changed{P-Boxes of the fourteen parameters updated in case II. Blue indicates the p-boxes of the parameters in the matrix $\bm{A}$, purple indicates parameters in the matrix $\bm{B}$.}}
    \label{fig:pboxes-case-II}
\end{figure}

\changed{\Cref{fig:pboxes-case-II} shows the p-boxes of all fourteen parameters updated in case II along with the respective prior ranges. It is evident that the updating reduces the ranges of parameter values quite substantially. Only for the last parameter, $\theta_{14} = b_4$, the p-box is still rather large.}

\subsubsection{Validation with time-dependent antibiotics}

As a final part, a validation setup is considered in order to see how robust the calibrated model parameters are to a changed setup. 
Specifically, a time-dependent application of antibiotics is considered in the validation case. As indicated for $\model^3_{\mathrm{val}}$ in \cref{tab:fixed-case-II}, antibiotics are applied $t=0.5$. For $t < 0.5$, the setup is identical with model $\model^3$ used to calibrate the final interaction parameters.

\begin{figure}[t]
    \centering
    \includegraphics[width=\linewidth]{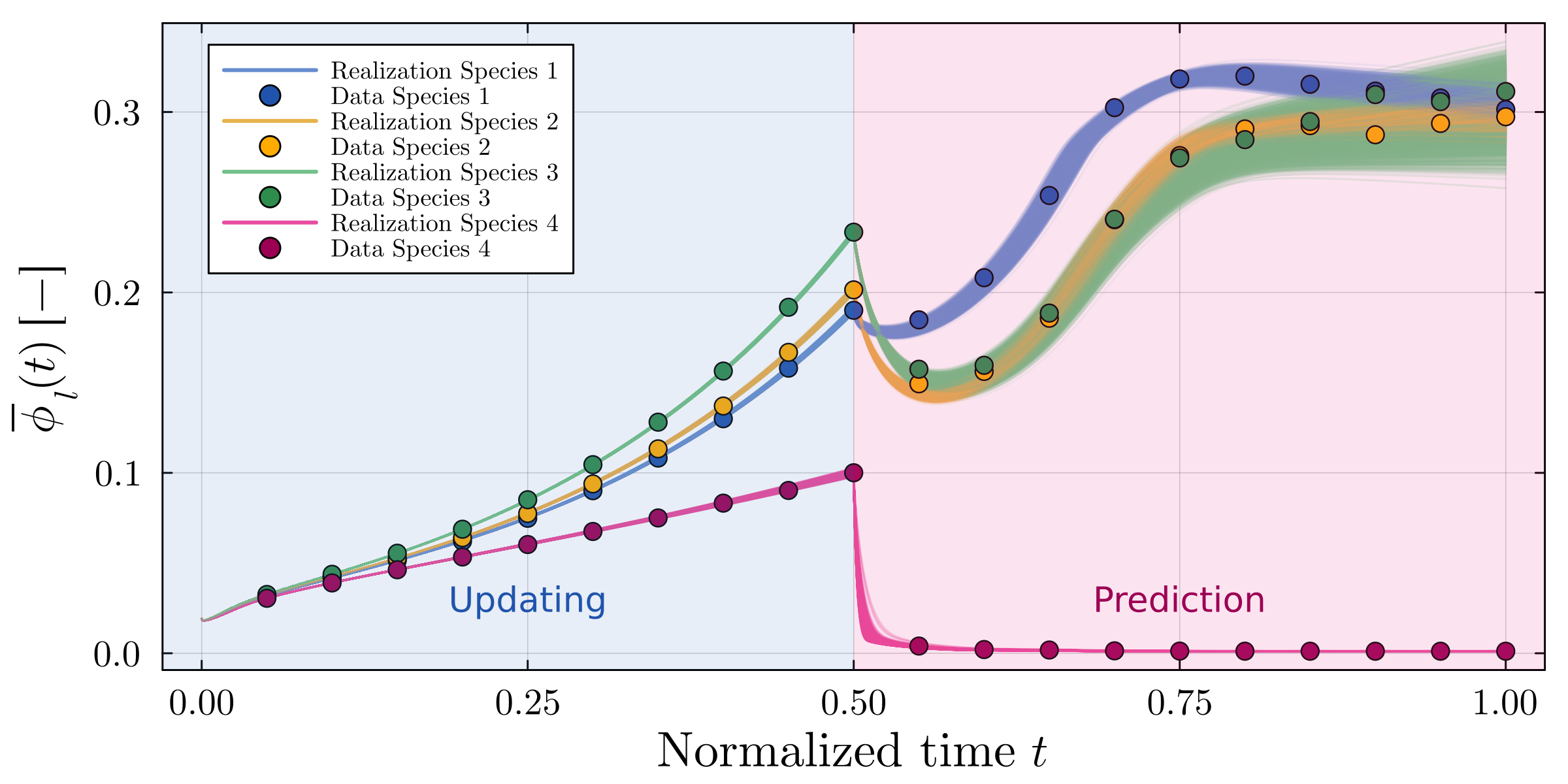}
    \caption{Comparison of the model output of model $\model^3_{\mathrm{val}}$ corresponding to calibrated posterior samples (shaded) and the data (scatter). The output for $t < 0.5$ corresponds to $\model^3$ from \cref{fig:output-M3}.}
    \label{fig:output-M3-val}
\end{figure}

The result of the validation is shown in \cref{fig:output-M3-val}. There, the model responses up until the antibiotics are applied are the same in for $\model^3$.
When the antibiotics are applied, a rapid change of the antibiotic concentration can be observed.
Again, we see a good agreement between the data and the predicted model response. Here, the data is only used to validate the predicted response by comparison; no additional model calibration is performed.
We note that the application of the antibiotics leads to an increased variability in the model response for the same set of inputs. This validation also shows that calibrating a physical model is useful in settings where the model is used in a predictive setting. Since the underlying model parameters were updated, a change in environmental conditions can be modeled, even if no data is available for this change.

\section{Conclusion}
\label{sec:conclusion}

In this paper, we presented a Bayesian updating approach for \changedIKM{the constitutive parameters of} biofilm growth models that accounts for hybrid uncertainties, incorporating both epistemic (unknown parameters) and aleatory (biological variability) uncertainty via a probabilistic formulation. Traditional double-loop approaches to uncertainty quantification are often inefficient in this context. To address this, we employed a reduced-order model based on \changed{TSM}, enabling the propagation of aleatory uncertainty with only a single model evaluation, thus eliminating the need for nested simulation loops. By leveraging a Taylor-decomposition-based representation of the stochastic process, our approach allows direct computation of the outputs mean and variance, which are then used in a Gaussian likelihood function for inference. This significantly reduces computational costs while preserving accuracy in capturing uncertainty effects compared to a Monte Carlo approach. 

The proposed methodology was validated through two representative case studies. The first involves a two-species biofilm model with five parameters, where monolithic updating was used to infer parameters governing growth, interaction, and antibiotic sensitivity. The second expands to a four-species system with fourteen parameters, employing a hierarchical inference strategy to decompose the high-dimensional problem into tractable sub-tasks. In both cases, the model successfully recovered the true parameters and revealed meaningful interdependencies among them, highlighting the ability of the method to capture complex inter-species dynamics. Additionally, a validation study using time-dependent antibiotic application confirmed that the inferred parameters retain predictive power under varying experimental conditions. This supports the robustness and generalizability of the calibrated model.

Furthermore, our results show that employing the TSM-ROM approach for Bayesian updating is robust to nonlinearities and suitable to deal with a large number of uncertain parameters. Moreover, since the TSM-ROM directly captures the output's uncertainties, the necessity for expensive multi-query simulations is removed and the updating is less demanding in terms of computation time. The proposed approach therefore has the ability to reduce the time it takes to infer model parameters. Thus, the TSM-ROM approach can be used to accelerate the verification of the presented biofilm model against \textit{in vitro} biofilm data, as for example presented by \cite{heine2025}.

Overall, the TSM-ROM approach offers a computationally efficient and robust framework for Bayesian inference in complex, uncertainty-laden biological systems. In future research, the computational demand can further be addressed by using more efficient updating schemes like variational inference \cite{Rubio2019RealtimeBayesian} or more efficient sampling strategies \cite{Igea2023CyclicalVariational}. 

\section*{Acknowledgements}
The work has been funded by the German Research Foundation (DFG) within the framework of the International Research Training Group IRTG 2657 ``Computational Mechanics Techniques in High Dimensions'' under grant number 433082294. 
The work has been funded by the European Union (ERC, Gen-TSM, project number 101124463). Views and opinions expressed are however those of the author(s) only and do not necessarily reflect those of the European Union or the European Research Council Executive Agency. Neither the European Union nor the granting authority can be held responsible for them.
The work has been funded by dtec.bw - Digitalization and Technology Research Center of the Bundeswehr. dtec.bw is funded by the European Union - NextGenerationEU.
The work has been funded by the German Research Foundation (Deutsche Forschungsgemeinschaft, DFG) through the project grant SFB/TRR-298-SIIRI  (Project-ID 426335750). 

\begin{appendices}

\section{Likelihood construction under hybrid uncertainties}\label{sec:likelihood}

In standard Bayesian updating frameworks (see, e.g., \cite{beck1998,lye2021}), the unknown parameters $\ftheta$ are typically regarded as purely epistemic quantities, with no consideration of aleatory uncertainty in the parameters and for that matter in the computational model. In this setting, model outputs are deterministic \cite{kitahara2022b}, and the only source of randomness is attributed to measurement error, $\bm{\varepsilon}$, as introduced in \cref{eq:model}.

When hybrid uncertainties are present, such as category IV parameters in \cref{fig:parameters}, this assumption no longer holds. The model itself becomes stochastic, meaning that uncertainty arises not only from measurement error $\bm{\varepsilon}$ but also as an \emph{inherent property} of the model \cite{kitahara2022b}. We therefore denote such models as stochastic models, $\hat{\model}$, which produce output sequences $\mathcal{Y}$ as defined in \cref{eq:stochastic_model}.

This shift has direct implications for the construction of the likelihood function. As we have defined earlier in \cref{sec:bmu}, the likelihood represents the probability of observing the data $\data$ for the parameter $\ftheta$.
In general, the data can be composed of $N$ observations and when it is assumed that these observations are independent, the likelihood is given as:

\begin{equation}
    \likelihoodcal = \likelihood = \prod_{k=1}^{N} p(\data^k| \ftheta).
\end{equation}

In the classical and purely epistemic case, the probability density function (PDF) $p(D^k | \ftheta)$ is given by the assumptions of the measurement error $\bm{\varepsilon}$. For the Gaussian assumption, for example, this PDF is given by a Gaussian PDF centered around the mean given my the model prediction $\bm{y} = \model(\ftheta)$ with covariance matrix $\bm{\Sigma}_{\bm{\varepsilon}}$:
\begin{equation}
    p(\data^k | \ftheta) = \frac{1}{\sqrt{(2 \pi)^{N_{\text{out}}} \det \bm{\Sigma}_{\bm{\varepsilon}}}} \exp \left[ - \frac{1}{2}(\data^k - \bm{y})^\top \bm{\Sigma}^{-1}_{\bm{\varepsilon}} (\data^k - \bm{y})\right] .
    \label{eq:GaussLkl}
\end{equation}

With hybrid uncertainties, however, the variability in the model outputs must also be taken into account in the PDF $p(\data^k| \ftheta)$. In practice, this requires estimating the aleatory output distribution of the stochastic model, conditional on fixed epistemic parameters \cite{bi2023}. Unfortunately, for complex engineering models, these output distributions cannot generally be predicted analytically from the input. Instead, they must be estimated. For instance, \cite{patelli2015,kitahara2022} propose kernel density estimation of the output PDF $p(\data^k| \ftheta)$, though this approach demands a very large number of forward model evaluations.

To mitigate this computational burden, numerical approximations of the likelihood function can be employed \cite{rocchetta2018a}. This family of approaches is commonly referred to as Approximate Bayesian Computation (ABC); an overview is provided in \cite{turner2012}. In ABC, the full likelihood is replaced by a kernel-based approximation, where a stochastic distance metric $d$ quantifies the discrepancy between the observed data $\data$ and the distribution of model outputs $\bm{y}$. \citet{kitahara2022b} gives a universal definition of the ABC likelihood as

\begin{equation}
    \likelihood \propto \frac{1}{\varepsilon} K\left\{ \frac{d(\bm{y}, \data)}{\varepsilon} \right\}.
\end{equation}

Here, $K$ presents the kernel function and $\varepsilon$ is a with factor. A common choice for the kernel is a squared exponential function \cite{bi2019a,bi2023,rocchetta2018a} which results in an approximated likelihood shaped similarly to a Gaussian PDF. $d(\bm{y}, \data)$ is a distance metric.
Various formulations for $d$ have been proposed and discussed in the literature, for an overview we refer to Refs. \cite{kitahara2022,bi2023,lye2024,bi2019a}.
\changed{In the context of this paper $d$ is a simple Gaussian since we observe a single data point per species and time step. Therefore the probability of observing the data under the respective model assumptions can directly be calculated.}

\end{appendices}

\newpage
\bibliography{biofilm.bib}

@article{beck1998,
  title = {Updating {{Models}} and {{Their Uncertainties}}. {{I}}: {{Bayesian Statistical Framework}}},
  shorttitle = {Updating {{Models}} and {{Their Uncertainties}}. {{I}}},
  author = {Beck, J. L. and Katafygiotis, L. S.},
  year = 1998,
  month = apr,
  journal = {J. Eng. Mech.},
  volume = {124},
  number = {4},
  pages = {455--461},
  issn = {0733-9399, 1943-7889},
  doi = {10.1061/(ASCE)0733-9399(1998)124:4(455)},
  urldate = {2024-05-04},
  langid = {english}
}

@article{beer2013,
  title = {Imprecise Probabilities in Engineering Analyses},
  author = {Beer, Michael and Ferson, Scott and Kreinovich, Vladik},
  year = 2013,
  month = may,
  journal = {Mech. Syst. Signal Process.},
  volume = {37},
  number = {1-2},
  pages = {4--29},
  issn = {08883270},
  doi = {10.1016/j.ymssp.2013.01.024},
  urldate = {2024-04-16},
  langid = {english}
}

@misc{behrensdorf2025,
  title = {{{FriesischScott}}/{{UncertaintyQuantification}}.{jl}: {{V0}}.12.0},
  shorttitle = {{{FriesischScott}}/{{UncertaintyQuantification}}.{jl}},
  author = {Behrensdorf, Jasper and Gray, Ander and Perin, Andrea and Grashorn, Jan and Luttmann, Max and Broggi, Matteo and Agarwal, Gopal and Fritsch, Lukas and Mett, Felix and Knipper, Laurenz},
  year = 2025,
  month = feb,
  doi = {10.5281/zenodo.14901342},
  urldate = {2025-02-20},
  howpublished = {Zenodo}
}

@article{betz2016,
  title = {Transitional {{Markov Chain Monte Carlo}}: {{Observations}} and {{Improvements}}},
  shorttitle = {Transitional {{Markov Chain Monte Carlo}}},
  author = {Betz, Wolfgang and Papaioannou, Iason and Straub, Daniel},
  year = 2016,
  month = may,
  journal = {J. Eng. Mech.},
  volume = {142},
  number = {5},
  pages = {04016016},
  issn = {0733-9399, 1943-7889},
  doi = {10.1061/(ASCE)EM.1943-7889.0001066},
  urldate = {2024-05-08},
  langid = {english}
}

@article{bi2019a,
  title = {The Role of the {{Bhattacharyya}} Distance in Stochastic Model Updating},
  author = {Bi, Sifeng and Broggi, Matteo and Beer, Michael},
  year = 2019,
  month = feb,
  journal = {Mech. Syst. Signal Process.},
  volume = {117},
  pages = {437--452},
  issn = {08883270},
  doi = {10.1016/j.ymssp.2018.08.017},
  urldate = {2024-04-29},
  langid = {english}
}

@article{bi2023,
  title = {Stochastic {{Model Updating}} with {{Uncertainty Quantification}}: {{An Overview}} and {{Tutorial}}},
  shorttitle = {Stochastic {{Model Updating}} with {{Uncertainty Quantification}}},
  author = {Bi, Sifeng and Beer, Michael and Cogan, Scott and Mottershead, John},
  year = 2023,
  month = dec,
  journal = {Mech. Syst. Signal Process.},
  volume = {204},
  pages = {110784},
  issn = {08883270},
  doi = {10.1016/j.ymssp.2023.110784},
  urldate = {2024-08-05},
  langid = {english}
}

@inproceedings{chamoin2025,
  title = {Data {{Assimilation}} and~{{Integration Into Computational Mechanics Models}}},
  booktitle = {Math. {{Comput}}. {{Model}}. {{Scales}}},
  author = {Chamoin, Ludovic},
  editor = {Diez, Pedro and Giacomini, Matteo},
  year = 2025,
  pages = {39--76},
  publisher = {Springer Nature Switzerland},
  address = {Cham},
  doi = {10.1007/978-3-031-84897-1_2},
  isbn = {978-3-031-84897-1},
  langid = {english}
}

@article{chattopadhyay2022,
  title = {Exploring the Role of Microbial Biofilm for Industrial Effluents Treatment},
  author = {Chattopadhyay, Indranil and J, Rajesh Banu and Usman, T. M. Mohamed and Varjani, Sunita},
  year = 2022,
  month = mar,
  journal = {Bioengineered},
  volume = {13},
  number = {3},
  pages = {6420--6440},
  publisher = {Taylor \& Francis},
  issn = {2165-5979},
  doi = {10.1080/21655979.2022.2044250},
  urldate = {2025-07-22},
  pmid = {35227160}
}

@article{ching2007,
  title = {Transitional {{Markov Chain Monte Carlo Method}} for {{Bayesian Model Updating}}, {{Model Class Selection}}, and {{Model Averaging}}},
  author = {Ching, Jianye and Chen, Yi-Chu},
  year = 2007,
  month = jul,
  journal = {J. Eng. Mech.},
  volume = {133},
  number = {7},
  pages = {816--832},
  issn = {0733-9399, 1943-7889},
  doi = {10.1061/(ASCE)0733-9399(2007)133:7(816)},
  urldate = {2024-05-03},
  langid = {english}
}

@phdthesis{dannert2023,
  title = {Numerical Treatment of Imprecise Random Fields in Non-Linear Solid Mechanics},
  author = {Dannert, Mona Madlen},
  year = 2023,
  url = {https://doi.org/10.15488/13241},
  langid = {english},
  school = {Leibniz University Hannover}
}

@article{donlan2002,
  title = {Biofilms: {{Microbial Life}} on {{Surfaces}}},
  author = {Donlan, Rodney},
  year = 2002,
  journal = {Emerg. Infect. Dis. J.},
  volume = {8},
  number = {9},
  pages = {881},
  issn = {1080-6059},
  doi = {10.3201/eid0809.020063}
}

@article{faes2021,
  title = {Engineering Analysis with Probability Boxes: {{A}} Review on Computational Methods},
  shorttitle = {Engineering Analysis with Probability Boxes},
  author = {Faes, Matthias and Daub, Marco and Marelli, Stefano and Patelli, Edoardo and Beer, Michael},
  year = 2021,
  month = nov,
  journal = {Struct. Saf.},
  volume = {93},
  pages = {102092},
  issn = {01674730},
  doi = {10.1016/j.strusafe.2021.102092},
  langid = {english}
}

@article{feng2021,
  title = {Modeling of {{Symbiotic Bacterial Biofilm Growth}} with an {{Example}} of the {{Streptococcus}}--{{Veillonella}} Sp. {{System}}},
  author = {Feng, Dianlei and Neuweiler, Insa and Nogueira, Regina and Nackenhorst, Udo},
  year = 2021,
  month = mar,
  journal = {Bull. Math. Biol.},
  volume = {83},
  number = {5},
  pages = {48},
  issn = {1522-9602},
  doi = {10.1007/s11538-021-00888-2},
  urldate = {2025-07-23},
  langid = {english}
}

@article{gabor2015,
  title = {Robust and Efficient Parameter Estimation in Dynamic Models of Biological Systems},
  author = {G{\'a}bor, Attila and Banga, Julio R.},
  year = 2015,
  month = oct,
  journal = {BMC Syst. Biol.},
  volume = {9},
  number = {1},
  pages = {74},
  issn = {1752-0509},
  doi = {10.1186/s12918-015-0219-2},
  urldate = {2025-07-21}
}

@article{geisler2023,
  title = {Time-Separated Stochastic Mechanics for the Simulation of Viscoelastic Structures with Local Random Material Fluctuations},
  author = {Geisler, Hendrik and Junker, Philipp},
  year = 2023,
  month = mar,
  journal = {Comput. Methods Appl. Mech. Eng.},
  volume = {407},
  pages = {115916},
  issn = {0045-7825},
  doi = {10.1016/j.cma.2023.115916},
  urldate = {2025-05-27}
}

@article{geisler2025,
  title = {A New Paradigm for the Efficient Inclusion of Stochasticity in Engineering Simulations: {{Time-separated}} Stochastic Mechanics},
  shorttitle = {A New Paradigm for the Efficient Inclusion of Stochasticity in Engineering Simulations},
  author = {Geisler, Hendrik and Erdogan, Cem and Nagel, Jan and Junker, Philipp},
  year = 2025,
  month = jan,
  journal = {Comput. Mech.},
  volume = {75},
  number = {1},
  pages = {211--235},
  issn = {1432-0924},
  doi = {10.1007/s00466-024-02500-5},
  urldate = {2025-01-10},
  langid = {english}
}

@article{hastings1970a,
  title = {Monte {{Carlo}} Sampling Methods Using {{Markov}} Chains and Their Applications},
  author = {Hastings, W. K.},
  year = 1970,
  month = apr,
  journal = {Biometrika},
  volume = {57},
  number = {1},
  pages = {97--109},
  issn = {0006-3444},
  doi = {10.1093/biomet/57.1.97}
}

@article{heine2025,
  title = {Influence of Species Composition and Cultivation Condition on Peri-Implant Biofilm Dysbiosis in Vitro},
  author = {Heine, Nils and Bittroff, Kristina and Szafra{\'n}ski, Szymon P. and Duitscher, Maya and Behrens, Wiebke and Vollmer, Clarissa and Mikolai, Carina and Kommerein, Nadine and Debener, Nicolas and Frings, Katharina and Heisterkamp, Alexander and Scheper, Thomas and {Torres-Mapa}, Maria L. and Bahnemann, Janina and Stiesch, Meike and {Doll-Nikutta}, Katharina},
  year = 2025,
  month = sep,
  journal = {Front. Oral Health},
  volume = {6},
  publisher = {Frontiers},
  issn = {2673-4842},
  doi = {10.3389/froh.2025.1649419},
  urldate = {2025-10-21},
  langid = {english}
}

@article{Igea2023CyclicalVariational,
  title = {Cyclical {{Variational Bayes Monte Carlo}} for Efficient Multi-Modal Posterior Distributions Evaluation},
  author = {Igea, Felipe and Cicirello, Alice},
  year = 2023,
  month = mar,
  journal = {Mech. Syst. Signal Process.},
  volume = {186},
  pages = {109868},
  issn = {0888-3270},
  doi = {10.1016/j.ymssp.2022.109868},
  urldate = {2024-10-01}
}

@article{james1995,
  title = {Interspecies Bacterial Interactions in Biofilms},
  author = {James, G A and Beaudette, L and Costerton, J W},
  year = 1995,
  month = oct,
  journal = {J. Ind. Microbiol.},
  volume = {15},
  number = {4},
  pages = {257--262},
  publisher = {Oxford University Press (OUP)},
  issn = {0169-4146, 1476-5535},
  doi = {10.1007/bf01569978},
  urldate = {2025-07-23},
  langid = {english}
}

@article{junker2021,
  title = {An Extended {{Hamilton}} Principle as Unifying Theory for Coupled Problems and Dissipative Microstructure Evolution},
  author = {Junker, Philipp and Balzani, Daniel},
  year = 2021,
  month = jul,
  journal = {Contin. Mech. Thermodyn.},
  volume = {33},
  number = {4},
  pages = {1931--1956},
  issn = {1432-0959},
  doi = {10.1007/s00161-021-01017-z}
}

@article{kang2023,
  title = {Strategies and Materials for the Prevention and Treatment of Biofilms},
  author = {Kang, Xiaoxia and Yang, Xiaoxiao and He, Yue and Guo, Conglin and Li, Yuechen and Ji, Haiwei and Qin, Yuling and Wu, Li},
  year = 2023,
  month = dec,
  journal = {Mater. Today Bio},
  volume = {23},
  pages = {100827},
  issn = {2590-0064},
  doi = {10.1016/j.mtbio.2023.100827},
  urldate = {2025-07-22}
}

@article{khatoon2018,
  title = {Bacterial Biofilm Formation on Implantable Devices and Approaches to Its Treatment and Prevention},
  author = {Khatoon, Zohra and McTiernan, Christopher D. and Suuronen, Erik J. and Mah, Thien-Fah and Alarcon, Emilio I.},
  year = 2018,
  month = dec,
  journal = {Heliyon},
  volume = {4},
  number = {12},
  pages = {e01067},
  issn = {2405-8440},
  doi = {10.1016/j.heliyon.2018.e01067},
  urldate = {2025-07-22}
}

@article{kirkpatrick1983,
  title = {Optimization by Simulated Annealing},
  author = {Kirkpatrick, S. and Gelatt, C. D. and Vecchi, M. P.},
  year = 1983,
  journal = {Science},
  volume = {220},
  number = {4598},
  pages = {671--680},
  doi = {10.1126/science.220.4598.671}
}

@article{kitahara2022,
  title = {Nonparametric {{Bayesian}} Stochastic Model Updating with Hybrid Uncertainties},
  author = {Kitahara, Masaru and Bi, Sifeng and Broggi, Matteo and Beer, Michael},
  year = 2022,
  month = jan,
  journal = {Mech. Syst. Signal Process.},
  volume = {163},
  pages = {108195},
  issn = {08883270},
  doi = {10.1016/j.ymssp.2021.108195},
  urldate = {2024-04-16},
  langid = {english}
}

@phdthesis{kitahara2022b,
  title = {Distribution-Free Stochastic Simulation Methodology for Model Updating under Hybrid Uncertainties},
  author = {Kitahara, Masaru},
  year = 2022,
  url = {https://doi.org/10.15488/12223},
  langid = {english},
  school = {Leibniz University Hannover}
}

@article{kiureghian2009,
  title = {Aleatory or Epistemic? {{Does}} It Matter?},
  shorttitle = {Aleatory or Epistemic?},
  author = {Kiureghian, Armen Der and Ditlevsen, Ove},
  year = 2009,
  month = mar,
  journal = {Struct. Saf.},
  volume = {31},
  number = {2},
  pages = {105--112},
  issn = {01674730},
  doi = {10.1016/j.strusafe.2008.06.020},
  urldate = {2024-04-16},
  langid = {english}
}

@article{klapper2010,
  title = {Mathematical {{Description}} of {{Microbial Biofilms}}},
  author = {Klapper, Isaac and Dockery, Jack},
  year = 2010,
  month = jan,
  journal = {SIAM Rev.},
  volume = {52},
  number = {2},
  pages = {221--265},
  publisher = {{Society for Industrial and Applied Mathematics}},
  issn = {0036-1445},
  doi = {10.1137/080739720},
  urldate = {2025-07-22}
}

@misc{klempt2025,
  title = {A Continuum Multi-Species Biofilm Model with a Novel Interaction Scheme},
  author = {Klempt, Felix and Geisler, Hendrik and Soleimani, Meisam and Junker, Philipp},
  year = 2025,
  eprint = {2509.01274},
  primaryclass = {cs.CE},
  url = {https://arxiv.org/abs/2509.01274},
  archiveprefix = {arXiv}
}

@article{kommerein2017,
  title = {An Oral Multispecies Biofilm Model for High Content Screening Applications},
  author = {Kommerein, Nadine and Stumpp, Sascha N. and M{\"u}sken, Mathias and Ehlert, Nina and Winkel, Andreas and H{\"a}ussler, Susanne and Behrens, Peter and Buettner, Falk F. R. and Stiesch, Meike},
  year = 2017,
  month = mar,
  journal = {PLOS ONE},
  volume = {12},
  number = {3},
  pages = {e0173973},
  publisher = {Public Library of Science},
  issn = {1932-6203},
  doi = {10.1371/journal.pone.0173973},
  urldate = {2025-07-23},
  langid = {english}
}

@article{lye2021,
  title = {Sampling Methods for Solving {{Bayesian}} Model Updating Problems: {{A}} Tutorial},
  shorttitle = {Sampling Methods for Solving {{Bayesian}} Model Updating Problems},
  author = {Lye, Adolphus and Cicirello, Alice and Patelli, Edoardo},
  year = 2021,
  month = oct,
  journal = {Mech. Syst. Signal Process.},
  volume = {159},
  pages = {107760},
  issn = {08883270},
  doi = {10.1016/j.ymssp.2021.107760},
  urldate = {2024-05-03},
  langid = {english}
}

@article{lye2024,
  title = {Comparison between {{Distance Functions}} for {{Approximate Bayesian Computation}} to {{Perform Stochastic Model Updating}} and {{Model Validation}} under {{Limited Data}}},
  author = {Lye, Adolphus and Ferson, Scott and Xiao, Sicong},
  year = 2024,
  month = jun,
  journal = {ASCE-ASME J Risk Uncertain. Eng Syst Part A: Civ Eng},
  volume = {10},
  number = {2},
  pages = {03124001},
  issn = {2376-7642},
  doi = {10.1061/AJRUA6.RUENG-1223},
  urldate = {2025-02-04},
  langid = {english}
}

@article{marsh2005,
  title = {Dental Plaque: {{Biological}} Significance of a Biofilm and Community Life-Style},
  shorttitle = {Dental Plaque},
  author = {Marsh, P. D.},
  year = 2005,
  journal = {J. Clin. Periodontol.},
  volume = {32},
  number = {s6},
  pages = {7--15},
  issn = {1600-051X},
  doi = {10.1111/j.1600-051X.2005.00790.x},
  urldate = {2025-07-23},
  langid = {english}
}

@incollection{mary-huard2011,
  title = {Introduction to {{Statistical Methods}} for {{Complex Systems}}},
  booktitle = {Handbook of Statistical Systems Biology},
  author = {{Mary-Huard}, Tristan and Robin, St{\'e}phane},
  editor = {Stumpf, Michael P. H. and Balding, David and Girolami, Mark},
  year = 2011,
  month = oct,
  edition = {First},
  pages = {15--38},
  publisher = {Wiley},
  address = {Chichester, West Sussex},
  doi = {10.1002/9781119970606.ch2},
  urldate = {2025-07-22},
  isbn = {978-0-470-71086-9 978-1-119-97060-6},
  langid = {english}
}

@article{melo1997,
  title = {Biofouling in Water Systems},
  author = {Melo, L. F. and Bott, T. R.},
  year = 1997,
  month = may,
  journal = {Exp. Therm. Fluid Sci.},
  volume = {14},
  number = {4},
  pages = {375--381},
  issn = {0894-1777},
  doi = {10.1016/S0894-1777(96)00139-2},
  urldate = {2025-07-22}
}

@article{metropolis1953,
  title = {Equation of {{State Calculations}} by {{Fast Computing Machines}}},
  author = {Metropolis, Nicholas and Rosenbluth, Arianna W. and Rosenbluth, Marshall N. and Teller, Augusta H. and Teller, Edward},
  year = 1953,
  month = jun,
  journal = {J. Chem. Phys.},
  volume = {21},
  number = {6},
  pages = {1087--1092},
  issn = {0021-9606, 1089-7690},
  doi = {10.1063/1.1699114},
  urldate = {2023-11-06},
  langid = {english}
}

@article{mitra2019,
  title = {Parameter Estimation and Uncertainty Quantification for Systems Biology Models},
  author = {Mitra, Eshan D. and Hlavacek, William S.},
  year = 2019,
  month = dec,
  journal = {Curr. Opin. Syst. Biol.},
  volume = {18},
  pages = {9--18},
  issn = {2452-3100},
  doi = {10.1016/j.coisb.2019.10.006},
  urldate = {2025-07-21}
}

@article{moons2009,
  title = {Bacterial Interactions in Biofilms},
  author = {Moons, Pieter and Michiels, Chris W. and Aertsen, Abram},
  year = 2009,
  month = aug,
  journal = {Crit. Rev. Microbiol.},
  volume = {35},
  number = {3},
  pages = {157--168},
  publisher = {Taylor \& Francis},
  issn = {1040-841X},
  doi = {10.1080/10408410902809431},
  urldate = {2025-07-22},
  pmid = {19624252}
}

@article{nadell2009,
  title = {The Sociobiology of Biofilms},
  author = {Nadell, Carey D. and Xavier, Joao B. and Foster, Kevin R.},
  year = 2009,
  month = jan,
  journal = {FEMS Microbiol. Rev.},
  volume = {33},
  number = {1},
  pages = {206--224},
  publisher = {Oxford University Press (OUP)},
  issn = {1574-6976},
  doi = {10.1111/j.1574-6976.2008.00150.x},
  urldate = {2025-07-22},
  langid = {english}
}

@article{nooranidoost2023,
  title = {Bayesian Estimation of {{Pseudomonas}} Aeruginosa Viscoelastic Properties Based on Creep Responses of Wild Type, Rugose, and Mucoid Variant Biofilms},
  author = {Nooranidoost, Mohammad and Cogan, N.G. and Stoodley, Paul and Gloag, Erin S. and Hussaini, M. Yousuff},
  year = 2023,
  month = dec,
  journal = {Biofilm},
  volume = {5},
  pages = {100133},
  issn = {25902075},
  doi = {10.1016/j.bioflm.2023.100133},
  urldate = {2025-02-12},
  langid = {english}
}

@article{ouidir2022,
  title = {Overview of Multi-Species Biofilms in Different Ecosystems: {{Wastewater}} Treatment, Soil and Oral Cavity},
  shorttitle = {Overview of Multi-Species Biofilms in Different Ecosystems},
  author = {Ouidir, Tassadit and Gabriel, Bruno and Nait Chabane, Yassine},
  year = 2022,
  month = may,
  journal = {J. Biotechnol.},
  volume = {350},
  pages = {67--74},
  issn = {0168-1656},
  doi = {10.1016/j.jbiotec.2022.03.014},
  urldate = {2025-07-22}
}

@article{paquette2006,
  title = {Risk {{Factors}} for {{Endosseous Dental Implant Failure}}},
  author = {Paquette, David W. and Brodala, Nadine and Williams, Ray C.},
  year = 2006,
  month = jul,
  journal = {Dent. Clin. North Am.},
  series = {Implantology},
  volume = {50},
  number = {3},
  pages = {361--374},
  issn = {0011-8532},
  doi = {10.1016/j.cden.2006.05.002},
  urldate = {2025-07-23}
}

@article{patelli2015,
  title = {Uncertainty {{Management}} in {{Multidisciplinary Design}} of {{Critical Safety Systems}}},
  author = {Patelli, Edoardo and Alvarez, Diego A. and Broggi, Matteo and Angelis, Marco De},
  year = 2015,
  month = jan,
  journal = {J. Aerosp. Inf. Syst.},
  volume = {12},
  number = {1},
  pages = {140--169},
  issn = {2327-3097},
  doi = {10.2514/1.I010273},
  urldate = {2025-08-27},
  langid = {english}
}

@article{rath2017,
  title = {Biofilm Formation by the Oral Pioneer Colonizer {{Streptococcus}} Gordonii: {{An}} Experimental and Numerical Study},
  shorttitle = {Biofilm Formation by the Oral Pioneer Colonizer {{Streptococcus}} Gordonii},
  author = {Rath, Henryke and Feng, Dianlei and Neuweiler, Insa and Stumpp, Nico S. and Nackenhorst, Udo and Stiesch, Meike},
  year = 2017,
  month = mar,
  journal = {FEMS Microbiol. Ecol.},
  volume = {93},
  number = {3},
  pages = {fix010},
  issn = {0168-6496},
  doi = {10.1093/femsec/fix010},
  urldate = {2025-07-23}
}

@article{read2020,
  title = {Strategies for Calibrating Models of Biology},
  author = {Read, Mark N and Alden, Kieran and Timmis, Jon and Andrews, Paul S},
  year = 2020,
  journal = {Brief. Bioinform.},
  publisher = {Oxford University Press (OUP)},
  issn = {1467-5463, 1477-4054},
  doi = {10.1093/bib/bby092},
  urldate = {2025-07-21},
  langid = {english}
}

@article{reiser2025,
  title = {Uncertainty Quantification and Propagation in Surrogate-Based {{Bayesian}} Inference},
  author = {Reiser, Philipp and Aguilar, Javier Enrique and Guthke, Anneli and B{\"u}rkner, Paul-Christian},
  year = 2025,
  month = mar,
  journal = {Stat Comput},
  volume = {35},
  number = {3},
  pages = {66},
  issn = {1573-1375},
  doi = {10.1007/s11222-025-10597-8},
  urldate = {2025-04-28},
  langid = {english}
}

@article{rittmann2018,
  title = {A Framework for Good Biofilm Reactor Modeling Practice ({{GBRMP}})},
  author = {Rittmann, Bruce E. and Boltz, Joshua P. and Brockmann, Doris and Daigger, Glen T. and Morgenroth, Eberhard and S{\o}rensen, Kim Hellesh{\o}j and Tak{\'a}cs, Imre and {van Loosdrecht}, Mark and Vanrolleghem, Peter A.},
  year = 2018,
  month = jan,
  journal = {Water Sci. Technol.},
  volume = {77},
  number = {5},
  pages = {1149--1164},
  issn = {0273-1223},
  doi = {10.2166/wst.2018.021},
  urldate = {2025-07-21}
}

@incollection{robert2011,
  title = {Bayesian {{Inference}} and {{Computation}}},
  booktitle = {Handbook of Statistical Systems Biology},
  author = {Robert, Christian P. and Marin, Jean-Michel and Rousseau, Judith},
  editor = {Stumpf, Michael P. H. and Balding, David and Girolami, Mark},
  year = 2011,
  month = oct,
  edition = {First},
  pages = {39--65},
  publisher = {Wiley},
  address = {Chichester, West Sussex},
  doi = {10.1002/9781119970606.ch3},
  urldate = {2025-07-22},
  isbn = {978-0-470-71086-9 978-1-119-97060-6},
  langid = {english}
}

@article{rocchetta2018a,
  title = {On-Line {{Bayesian}} Model Updating for Structural Health Monitoring},
  author = {Rocchetta, Roberto and Broggi, Matteo and Huchet, Quentin and Patelli, Edoardo},
  year = 2018,
  month = mar,
  journal = {Mech. Syst. Signal Process.},
  volume = {103},
  pages = {174--195},
  issn = {0888-3270},
  doi = {10.1016/j.ymssp.2017.10.015}
}

@article{Rubio2019RealtimeBayesian,
  title = {Real-Time {{Bayesian}} Data Assimilation with Data Selection, Correction of Model Bias, and on-the-{{Fly}} Uncertainty Propagation},
  author = {Rubio, Paul-Baptiste and Chamoin, Ludovic and Louf, Fran{\c c}ois},
  year = 2019,
  month = nov,
  journal = {Comptes Rendus M\'ecanique},
  series = {Data-{{Based Engineering Science}} and {{Technology}}},
  volume = {347},
  number = {11},
  pages = {762--779},
  issn = {1631-0721},
  doi = {10.1016/j.crme.2019.11.004},
  urldate = {2023-01-27},
  langid = {english}
}

@article{shewa2024,
  title = {Biofilm Characterization and Dynamic Simulation of Advanced Rope Media Reactor for the Treatment of Primary Effluent},
  author = {Shewa, Wudneh A. and Sun, Lin and Bossy, Kevin and Dagnew, Martha},
  year = 2024,
  journal = {Water Environ. Res.},
  volume = {96},
  number = {11},
  pages = {e11150},
  issn = {1554-7531},
  doi = {10.1002/wer.11150},
  urldate = {2025-07-21},
  copyright = {\copyright{} 2024 The Author(s). Water Environment Research published by Wiley Periodicals LLC on behalf of Water Environment Federation.},
  langid = {english}
}

@article{shree2023,
  title = {Biofilms: {{Understanding}} the Structure and Contribution towards Bacterial Resistance in Antibiotics},
  shorttitle = {Biofilms},
  author = {Shree, Pallee and Singh, Chandra Kant and Sodhi, Kushneet Kaur and Surya, Jaya Niranjane and Singh, Dileep Kumar},
  year = 2023,
  month = jun,
  journal = {Med. Microecol.},
  volume = {16},
  pages = {100084},
  issn = {2590-0978},
  doi = {10.1016/j.medmic.2023.100084},
  urldate = {2025-07-22}
}

@article{stefanou2009,
  title = {The Stochastic Finite Element Method: {{Past}}, Present and Future},
  shorttitle = {The Stochastic Finite Element Method},
  author = {Stefanou, George},
  year = 2009,
  month = feb,
  journal = {Comput. Methods Appl. Mech. Eng.},
  volume = {198},
  number = {9-12},
  pages = {1031--1051},
  issn = {00457825},
  doi = {10.1016/j.cma.2008.11.007},
  urldate = {2024-04-04},
}

@article{taghizadeh2020,
  title = {Bayesian Inversion for a Biofilm Model Including Quorum Sensing},
  author = {Taghizadeh, Leila and Karimi, Ahmad and Presterl, Elisabeth and Heitzinger, Clemens},
  year = 2020,
  month = feb,
  journal = {Comput. Biol. Med.},
  volume = {117},
  pages = {103582},
  issn = {0010-4825},
  doi = {10.1016/j.compbiomed.2019.103582},
  urldate = {2025-02-12}
}

@article{turner2012,
  title = {A Tutorial on Approximate {{Bayesian}} Computation},
  author = {Turner, Brandon M. and Van Zandt, Trisha},
  year = 2012,
  month = apr,
  journal = {J. Math. Psychol.},
  volume = {56},
  number = {2},
  pages = {69--85},
  issn = {00222496},
  doi = {10.1016/j.jmp.2012.02.005},
  urldate = {2024-08-05},
  langid = {english}
}

@article{wilkinson2007,
  title = {Bayesian Methods in Bioinformatics and Computational Systems Biology},
  author = {Wilkinson, Darren J.},
  year = 2007,
  month = mar,
  journal = {Brief Bioinform},
  volume = {8},
  number = {2},
  pages = {109--116},
  publisher = {Oxford Academic},
  issn = {1467-5463},
  doi = {10.1093/bib/bbm007},
  urldate = {2025-02-14},
  langid = {english}
}

@article{willmann2022,
  title = {Inverse Analysis of Material Parameters in Coupled Multi-Physics Biofilm Models},
  author = {Willmann, Harald and Wall, Wolfgang A.},
  year = 2022,
  month = jun,
  journal = {Adv. Model. Simul. Eng. Sci.},
  volume = {9},
  number = {1},
  pages = {7},
  issn = {2213-7467},
  doi = {10.1186/s40323-022-00220-0},
  urldate = {2025-08-07}
}

@article{willmann2022a,
  title = {Bayesian Calibration of Coupled Computational Mechanics Models under Uncertainty Based on Interface Deformation},
  author = {Willmann, Harald and Nitzler, Jonas and Brandst{\"a}ter, Sebastian and Wall, Wolfgang A.},
  year = 2022,
  month = dec,
  journal = {Adv. Model. Simul. Eng. Sci.},
  volume = {9},
  number = {1},
  pages = {24},
  issn = {2213-7467},
  doi = {10.1186/s40323-022-00237-5},
  urldate = {2025-08-07}
}

@article{wollner2025,
  title = {A Reparameterization-Invariant {{Bayesian}} Framework for Uncertainty Estimation and Calibration of Simple Materials},
  author = {Wollner, Maximilian P. and {Rolf-Pissarczyk}, Malte and Holzapfel, Gerhard A.},
  year = 2025,
  month = feb,
  journal = {Comput. Mech.},
  issn = {0178-7675, 1432-0924},
  doi = {10.1007/s00466-024-02573-2},
  urldate = {2025-08-07}
}

@article{yang2011,
  title = {Current Understanding of Multi-Species Biofilms},
  author = {Yang, Liang and Liu, Yang and Wu, Hong and H{\o}iby, Niels and Molin, S{\o}ren and Song, Zhi-jun},
  year = 2011,
  month = apr,
  journal = {Int. J. Oral Sci.},
  volume = {3},
  number = {2},
  pages = {74--81},
  publisher = {Nature Publishing Group},
  issn = {2049-3169},
  doi = {10.4248/IJOS11027},
  urldate = {2025-07-23},
  copyright = {2011 West China School of Stomatology},
  langid = {english}
}

@article{behmaneshHierarchicalBayesianModel2015,
  title = {Hierarchical {{Bayesian}} Model Updating for Structural Identification},
  author = {Behmanesh, Iman and Moaveni, Babak and Lombaert, Geert and Papadimitriou, Costas},
  year = 2015,
  month = dec,
  journal = {Mech. Syst. Signal Process.},
  volume = {64--65},
  pages = {360--376},
  issn = {08883270},
  doi = {10.1016/j.ymssp.2015.03.026},
  urldate = {2025-12-08}
}

\end{document}